\documentclass[utf8]{FrontiersinHarvard} 

\usepackage{url,hyperref,lineno,microtype,subcaption}
\usepackage[onehalfspacing]{setspace}
\usepackage{comment}

\newcommand{\chandra}{\textit{Chandra}}

\newcommand{\xmm}{\textit{XMM}}

\newcommand{\fermi}{\textit{Fermi}}
\newcommand{\swift}{\textit{Swift}}
\newcommand{\integral}{\textit{INTEGRAL}}
\newcommand{\nustar}{\textit{NuSTAR}}

\newcommand\hexp{\textit{HEX-P}}

\newcommand{\fluxcgs}{\ensuremath{\mathrm{erg}\,\mathrm{s}^{-1}\,\mathrm{cm}^{-2}}}

\def\amin{\ifmmode^{\prime}\else$^{\prime}$\fi}
\def\asec{\ifmmode^{\prime\prime}\else$^{\prime\prime}$\fi}
\def\simgt{\lower.5ex\hbox{$\; \buildrel > \over \sim \;$}}
\def\simlt{\lower.5ex\hbox{$\; \buildrel < \over \sim \;$}}

\def\keyFont{\fontsize{8}{11}\helveticabold }
\def\firstAuthorLast{Reynolds {et~al.}} 
\def\Authors{Stephen Reynolds\,$^{1}$, Hongjun An\,$^{2}$, Moaz Abdelmaguid\,$^{3}$, Jason Alford\,$^{3}$,  Chris L. Fryer\,$^{4}$, Kaya Mori\,$^{5, *}$, Melania Nynka\,$^{6}$, Jaegeun Park\,$^{2}$, Yukikatsu Terada\,$^{7,8}$, Jooyun Woo\,$^{5}$, Aya Bamba\,$^{9,10,11}$, Priyadarshini Bangale\,$^{12}$, Rebecca Diesing\,$^{13}$, Jordan Eagle\,$^{14}$, 
Stefano Gabici\,$^{15}$, Joseph Gelfand\,$^{3}$, Brian Grefenstette\,$^{16}$, Javier Garcia\,$^{14}$, Chanho Kim\,$^{2}$, Sajan Kumar\,$^{17}$, 
Brydyn Mac Intyre\,$^{18}$, Kristin Madsen\,$^{14}$, 
Silvia Manconi\,$^{19}$, Yugo Motogami\,$^{7}$, Hayato Ohsumi\,$^{7}$, Barbara Olmi\,$^{20,21}$, Toshiki Sato\,$^{22}$, Ruo-Yu Shang\,$^{23}$, Daniel Stern\,$^{24}$,  Naomi Tsuji\,$^{25}$, George Younes\,$^{14}$, and Andreas Zoglauer\,$^{26}$}
\def\Address{\footnotesize 
$^{1}$ Department of Physics, North Carolina State University, Raleigh, NC, USA\\  
$^{2}$ Department of Astronomy and Space Science, Chungbuk National University, Cheongju, Republic of Korea\\  
$^{3}$ Department of Physics, New York University, Abu Dhabi, UAE\\  
$^{4}$ Center for Non Linear Studies, Los Alamos National Laboratory, Los Alamos, NM, USA\\  
$^{5}$ Columbia Astrophysics Laboratory, Columbia University, New York, NY, USA\\
$^{6}$ Kavli Institute For Astrophysics and Space Research, Massachusetts Institute of Technology, Cambridge, MA, USA \\ 
$^{7}$ Graduate School of Science and Engineering, Saitama University, Sakura, Saitama, Japan\\ 
$^{8}$ Japan Aerospace Exploration Agency(JAXA), Institute of Space and Astronautical Science, Sagamihara, Japan\\
$^{9}$ Department of Physics, The University of Tokyo, Tokyo, Japan\\  
$^{10}$ Research Center for the Early Universe, School of Science, The University of Tokyo, Tokyo, Japan\\
$^{11}$ Trans-Scale Quantum Science Institute, The University of Tokyo, Tokyo, Japan\\
$^{12}$ Department of Physics and Astronomy and the Bartol Research Institute, University of Delaware, Newark, DE, USA\\  
$^{13}$ Department of Astronomy and Astrophysics, The University of Chicago, Chicago, IL, USA\\  
$^{14}$  NASA Goddard Space Flight Center, Greenbelt, MD, USA\\  
$^{15}$ Universite Paris Cite, CNRS, Astroparticule et Cosmologie, Paris, France\\  
$^{16}$ Space Radiation Laboratory, California Institute of Technology,  Pasadena, CA USA\\  
$^{17}$ Department of Physics, University of Maryland, College Park, MD 20742-4111, USA\\ 
$^{18}$ Department of Physics and Astronomy, University of Manitoba, Winnipeg, Canada\\  
$^{19}$ Laboratoire d’Annecy-le-Vieux de Physique Theorique, CNRS, USMB, Annecy, France\\  
$^{20}$ INAF - Osservatorio Astrofisico di Arcetri, Largo E. Fermi 5, I-50125 Firenze, Italy\\
$^{21}$ INAF - Osservatorio Astronomico di Palermo, Piazza del Parlamento 1, I-90134 Palermo, Italy\\
$^{22}$ Department of Physics, School of Science and Technology, Meiji University, Kanagawa, Japan\\  
$^{23}$ Department of Physics and Astronomy, Barnard College, New York, NY, USA\\  
$^{24}$ Jet Propulsion Laboratory, California Institute of Technology, Pasadena, CA, 91109, USA\\  
$^{25}$ Faculty of Science, Kanagawa University, Kanagawa, Japan\\  
$^{26}$ Space Sciences Laboratory, UC Berkeley, Berkeley, CA, USA  
}
\def\corrAuthor{\footnotesize Kaya Mori}
\def\corrEmail{\footnotesize kaya@astro.columbia.edu}

\begin{document}
\onecolumn
\firstpage{1}

\title {The High Energy X-ray Probe (HEX-P): Supernova remnants, pulsar wind nebulae, and nuclear astrophysics} 

\author[\firstAuthorLast ]{\Authors} 
\address{} 
\correspondance{} 

\extraAuth{}

\maketitle

\begin{abstract}

\section{}

HEX-P is a probe-class mission concept that will combine high spatial resolution X-ray imaging ($<10''$ full width at half maximum) and broad spectral coverage (0.2--80 keV) with an effective area far superior to current facilities (including \xmm-Newton and \nustar) to enable revolutionary new insights into a variety of important astrophysical problems. HEX-P is ideally suited to address important problems in the physics and astrophysics of supernova remnants (SNRs) and pulsar-wind nebulae (PWNe). For shell SNRs, HEX-P can greatly improve our understanding via more accurate spectral characterization and localization of non-thermal X-ray emission from both non-thermal-dominated SNRs and those containing both thermal and non-thermal components, and can discover previously unknown non-thermal components in SNRs.  Multi-epoch HEX-P observations of several young SNRs (e.g., Cas~A and Tycho) are expected to detect year-scale variabilities of X-ray filaments and knots, thus enabling us to determine fundamental parameters related to diffusive shock acceleration, such as local magnetic field strengths and maximum electron energies. 
For PWNe, HEX-P will provide spatially-resolved, broadband X-ray spectral data separately from their pulsar emission, allowing us to study how particle acceleration, cooling, and propagation operate in different evolution stages of PWNe. 
HEX-P is also poised to make unique and significant contributions to nuclear astrophysics of Galactic radioactive sources by improving detections of, or limits on, $^{44}$Ti in the youngest SNRs and by potentially discovering rare nuclear lines as evidence of double neutron star mergers. 
Throughout the paper, we present simulations of each class of objects, demonstrating the power of both the imaging and spectral capabilities of HEX-P to advance our knowledge of SNRs, PWNe, and nuclear astrophysics. 


\tiny
 \keyFont{ \section{Keywords:} supernova remnants, pulsar wind nebulae, nuclear astrophysics, X-ray astrophysics, future missions} 
\end{abstract}



\section{Introduction}

In the Galactic ecosystem, fast particles play a crucial role in ionization of molecular material (influencing star and planet formation), in regulating the interstellar magnetic field, and perhaps in driving a Galactic wind.  We observe a highly filtered subset of
those particles as cosmic rays at Earth, but their origins lie in a range of high-energy-density Galactic environments.  Strong shock waves from supernova remnants (SNRs) remain the prime candidate for accelerating the bulk of Galactic cosmic-ray electrons and ions up to energies of the order of PeV, through the diffusive shock acceleration (DSA) mechanism.  An additional leptonic component (electrons and positrons) is thought to originate in pulsar-wind nebulae (PWNe), 
probably at the termination shock of the initially relativistic cold outflow.  These classes of object exhibit the physics of particle acceleration in particularly accessible ways, through spatially resolved nonthermal emission in radio, X-ray, and gamma-ray bands.  That emission is due to some combination of four important processes, three involving leptons and one due to hadrons.   The leptonic processes, described in more detail in Mori et al.~(2023), are synchrotron radiation producing radio through hard X-ray emission,
inverse-Compton upscattering of ambient photon fields producing GeV to TeV gamma rays, and nonthermal bremsstrahlung producing hard X-rays through gamma rays.  Of these, the latter is rarely dominant.  The hadronic process begins with inelastic collisions of energetic ions with ambient thermal gas, producing charged and neutral pions (once the kinematic threshold of about 70 MeV is surpassed).  The charged pions decay, ultimately, to secondary electrons and positrons which can produce synchrotron radiation, in an inescapable consequence of the hadronic process.  The neutral pions decay to gamma rays, which can range in energy up to ten percent or so of that of the initiating 
hadron.  One of the fundamental questions in interpreting potential Galactic cosmic-ray sources involves deciding which process is responsible for their gamma-ray emission.

This paper demonstrates the potential contributions to understanding
these issues that can be provided by a high resolution, high sensitivity hard X-ray imaging capability. In addition to access to the continuum processes described above, the hard X-ray spectral region offers a window into some of the most energetic stellar-scale processes
in Nature: along with supernovae and SNRs, the kilonovae
resulting from merging binary neutron stars.  In both these arenas,
nucleosynthesis is important and can be diagnosed through the production of unstable species with a range of half-lives which decay with the production of different hard X-ray lines.  Of particular
relevance for SNe and young SNRs are the 68 and 78 keV lines
produced in the decay of $^{44}$Ti, but merging neutron stars are
also expected to produce potentially observable lines for a sufficiently nearby event.  Information from such decays gives
insight into nucleosynthesis unavailable in any other channel.

The HEX-P instrument can bring to the study of these issues  unmatched sensitivity and angular resolution above 10 keV.  Below, we lay out the contributions HEX-P can make to characterizing the nature and evolution of SNRs and PWNe. \S2 describes the current telescope design and HEX-P simulation tools. \S3 demonstrates HEX-P's spectro-imaging capabilities for characterizing thermal and non-thermal X-ray emission from young SNRs by showing simulation results for Cas A, Tycho, G1.9+0.3, and SN1987A. \S4 presents HEX-P's observation program of investigating different evolutionary stages of PWNe and exploring this primary class of leptonic particle accelerators in our Galaxy. HEX-P's broad-band X-ray data, in conjunction with the upcoming Cherenkov Telescope Array (CTA) mission, will allow us to dissect the particle acceleration, propagation, and cooling processes in PWNe.
\S5 discusses HEX-P's unique role in nuclear astrophysics, such as detecting $^{44}$Ti lines from the youngest SNRs and nuclear lines from double neutron star mergers in our Galaxy. In our companion paper (Mori et al., submitted to FrASS), we will discuss other types of Galactic particle accelerators such as unidentified PeVatron candidates, star clusters, superbubbles, microquasar jets, TeV gamma-ray binaries and the supermassive black hole in Sgr A*.  The science we describe here has considerable overlap with that discussed in parallel HEX-P papers.  Our planned studies of PWNe will of course also produce important new information on the pulsars they contain (see Jaodand et al., submitted to FrASS), with possible bearing on magnetars (Alford et al., submitted to FrASS) and general properties of neutron stars (Ludlum et al., submitted to FrASS) as well. Various classes of binary X-ray source are also described in Connors et al.~(submitted to FrASS).  Several of our
target sources may be observed as part of the Galactic Center effort (Mori et al., submitted to FrAss), while others lie in nearby galaxies (Lehmer et al., submitted to FrASS).  SN1987A is on our list of primary targets, and we describe here some of the nuclear astrophysics possible with SNe or neutron-star mergers, more likely to be found among nearby galaxies than in our own.

\section{HEX-P mission design and simulation}

The High-Energy X-ray Probe (HEX-P; Madsen+23) is a probe-class mission concept that offers sensitive broad-band coverage ($0.2-80$\,keV) of the X-ray spectrum with exceptional spectral, timing, and angular capabilities. It features two high-energy telescopes (HET) that focus hard X-rays, and a low-energy telescope (LET) providing soft X-ray coverage.

The LET consists of a segmented mirror assembly coated with Ir on monocrystalline silicon that achieves a half power diameter of $3.5''$, and a low-energy DEPFET detector, of the same type as the Wide Field Imager (WFI; Meidinger et al. 2020) onboard Athena (Nandra et al., 2013). It has 512 x 512 pixels that cover a field of view of $11.3' \times 11.3'$. It has an effective bandpass of $0.2-25$\,keV and a full-frame readout time of 2\,ms, and can be operated in 128 and 64 channel window modes for higher count-rates to mitigate pile-up and allow faster readout. Pile-up effects remain below an acceptable limit of $\sim 1$\% for sources up to $\sim 100$\,mCrab in the smallest window configuration (64w). Excising the core of the point-spread function (PSF), a common practice in X-ray astronomy, will allow for observations of brighter sources, with a maximum loss of $\sim 60\%$ of the total photon counts.

The HET consists of two co-aligned telescopes and detector modules. The optics are made of Ni-electroformed full shell mirror substrates, leveraging the heritage of \xmm\  \citep{Jansen2001}, and coated with Pt/C and W/Si multilayers for an effective bandpass of $2-80$\,keV. The high-energy detectors are of the same type as those flown on \nustar\ \citep{Harrison2013}, and they consist of 16 CZT sensors per focal plane, tiled $4 \times 4$, for a total of $128 \times 128$ pixel spanning a field of view slightly larger than for the LET, of $13.4' \times 13.4'$.

All the simulations presented here were produced with a set of response files that represent the observatory performance based on current best estimates (see Madsen et al., 2023). The effective area is derived from a ray-trace of the mirror design including obscuration by all known structures. The detector responses are based on simulations performed by the respective hardware groups, with an optical blocking filter for the LET and a Be window and thermal insulation for the HET. The LET background was derived from a GEANT4 simulation \citep{Eraerds2021} of the WFI instrument, and the one for the HET from a GEANT4 simulation of the \nustar\ instrument, both positioned at L1. Throughout the paper, we present our simulation results for \hexp\ using the SIXTE \citep{Dauser2019} and XSPEC toolkits \citep{Arnaud1996}. To ensure the most realistic simulation results, we incorporated recent high-resolution X-ray images (mostly from \chandra\ or other wavelength observations), the best-known  spectral information, and theoretical model predictions. Various exposure times have been considered for the feasibility studies presented in the following sections.

\section{Supernova remnants}

The best understood of all sources of high-energy particles are
probably the shell SNRs, where independent measures of the ambient
densities, shock speeds, and magnetic-field strengths are all
possible.  However, the evidence for those high-energy particles is
primarily synchrotron X-ray emission, unambiguously separable from
thermal emission only at energies above 10 keV or so.  \nustar\ showed
the promise of studying particle acceleration in SNRs with a few observations of young bright SNRs with strong nonthermal emission relative to thermal, but those observations raised more questions than they answered.  HEX-P will be able to identify and characterize hard nonthermal X-ray emission in many more remnants, and will be able to address many of the questions raised by earlier observations.  Older SNRs with somewhat slower shock velocities, predicted by theory to accelerate electrons to X-ray-emitting energies but too faint for prior study, can be examined by HEX-P, along with potential new sources turned up by upcoming surveys such as that performed by eROSITA, and newly occurring transients that may appear. Below we illustrate the capabilities of HEX-P with simulations of Cas A, Tycho, G1.9+0.3 (youngest Galactic SNR), and SN1987A.

\subsection{Thermal emission}

Supernova remnants in X-rays are chiefly thermal sources, with
temperatures from a few tenths to a few keV, and rich line spectra
providing essential information on plasma composition, and given sufficient energy resolution, on radial velocities and ionization state.  See \cite{vink12} for a review.  Most
young remnants (estimated ages less than a few thousand years) also
show spatial and spectral evidence for nonthermal X-rays, often in the
form of thin rims at the remnant periphery, and in hard continuum
components sometimes apparent in relatively line-free regions of the
spectrum between 4 and 6 keV. Some of the continuum emission in this
spectral region is certainly thermal bremsstrahlung, and interpreting the Fe K$\alpha$ line complex at 6.4 -- 6.7 keV requires knowing the thermal fraction. The nonthermal component can be followed to higher energies. Non-imaging instruments before \chandra\ and \xmm\ found hard continua in the integrated spectra of a few remnants, the most spectacular being Cas A, with emission extending to at least 100 keV \citep{vink00}. But few remnants are bright enough to be studied in this way.  \nustar\ was able to detect non-thermal continua in a few young remnants, to 20 keV or (in a few cases) 30 keV.  But in others, where small features in Chandra images are found to be line-free, the integrated emission at energies above 10 keV is not strong enough to be examined with \nustar.

The ability to identify thermal and nonthermal emission components in
SNRs below about 10 keV would provide a powerful tool for diagnosing
the detailed relation between particle acceleration and the properties
of the thermal plasma.  HEX-P's two instruments will provide such a
tool.  One simple but critical distinction would be to differentiateshocked ejecta from shocked surrounding material (undisturbed interstellar medium [ISM], for
remnants of SNe Ia, or modified stellar wind or circumstellar material
for remnants of core-collapse SNe), requiring good spatial and spectral
resolution in the line-rich region from 0.5 to 4 keV, as well as at the
crucial Fe K$\alpha$ line.  While it is possible to piece
together some of this information by combining \chandra\ or \xmm\ 
observations with, say, \nustar\ data, the task is complex in practice.
Having two instruments aboard the same spacecraft will significantly
improve our ability to cross-calibrate and will result in unbroken
spectra from as low energies as interstellar absorption typically
permits ($\simgt 0.2$ keV) up to 80 keV if such photons are present.
SNR plasmas are generally multitemperature, and while they can sometimes
be fairly well represented by two components, a higher-temperature component is easy to confuse with nonthermal emission, with important
effects on interpreting, say, the equivalent width of the Fe K$\alpha$
line complex.  In fact, the LET instrument alone, coupled with HEX-P's
large effective area, will improve the purely thermal diagnosis of
many SNRs.  Particle backgrounds constitute significant impediments
above 8 or 9 keV for both \chandra\ and \xmm, but can be expected
to be significantly lower at the L1 position of HEX-P. 

\subsection{Nonthermal continuum
}
  The presence in most young SNRs of strong thermal X-ray
emission is both a bug and a feature: below 10 keV, the fraction of
nonthermal emission is often quite uncertain, but as described above, thermal emission provides essential information on densities and shock speeds, and composition.  
Observing those objects at X-ray energies above 10 keV where the
thermal contribution is small or negligible then allows the separate
characterization of the nonthermal-electron energy spectrum.  Basic questions afflict our understanding of electron acceleration to
high energies in young SNRs.  Where, exactly, does the acceleration
take place?  What are the physical conditions there, as indicated by
observations at other wavelengths?  What are the spectral shapes?  To
how high photon energies does emission extend, and are there spatial
variations in spectral shape?  The current fleet of X-ray observatories has produced important information on some of these
questions, but the acceleration to the highest energies remains
mysterious.  Only \nustar\ has been able to image SNRs well above
thermal energies, and spatial resolution, effective area, and
background limitations have confined those studies to a few objects.
But what has been learned is alarming: especially in Cas A, both
spectral and spatial expectations for the highest-energy synchrotron
emission were impressively upended.  Neither the forward nor the
reverse shock seems to be the site of the strongest emission between 15
and 50 keV.  The spectra, while steeper than radio, show no indication
of further curvature, indicating that the highest-energy electrons
have not yet been observed.  Observations of the handful of other young SNRs bright enough above 10 keV for \nustar\ hint at similar deficiencies.

But this problem is an opportunity.  The study of particle
acceleration in shocks has seemed to be an essentially solved problem,
with the only remaining tasks to measure the parameters of the
standard models. But those standard models are clearly incomplete, in 
the face of the \nustar\ data.  More detailed observations above 10 keV
of not only the historical remnants but other young objects less
well-known, stand to produce qualitatively new information most 
relevant to basic questions of efficiency and maximum energies
attainable (e.g., see \citet{Diesing2021}) that may apply to acceleration sites elsewhere in the
Universe.

\subsection{Cassiopeia A}

Cassiopeia A (Cas A) is the youngest remnant of a core-collapse supernova event in our Galaxy. Light-echo observations of the
scattered optical emission from the supernova event \citep{2008Sci...320.1195K} have established the supernova subtype as IIb -- the explosion of a massive star that lost
most of its hydrogen envelope, and has expanded into the pre-SN stellar wind. 
Its radio emission is dramatically stronger than that
from any other Galactic SNR, perhaps partly due to the blast wave
having encountered dense stellar wind at earlier times.  The strongest
radio emission from electrons with GeV energies is found not at
the outer blast wave but at a bright inner ring between the forward
and reverse shocks, where strong magnetic turbulence is likely to be present, boosting synchrotron emissivity because of a stronger magnetic field but also perhaps further accelerating nonthermal particles.  
These properties distinguish Cas A from young remnants of Type Ia
supernovae such as Tycho and Kepler, which expand into inhomogeneous,
but probably not radially varying, interstellar medium, and whose
radio emission tends to be concentrated at the forward shock.  

While the bright radio ring in Cas A indicates a large population
of GeV-energy electrons, it was expected that the highest-energy X-ray-emitting particles would
be found at the primary acceleration sites in the remnant, probably
the forward blast wave.  Evidence for particle acceleration at reverse shocks in young SNRs is ambiguous at best.  The \nustar\ observations \citep{NuSTAR_CasA15}, however, told a different story,
showing diffuse hard
X-ray ($>$ 15 keV) emission from the remnant interior, with the brightest emission from two compact knots inside the blast wave
to the west. Hard emission is present and associated with the outer
blast wave, but it is considerably fainter.  There are approximate
correspondences between the bright knots and features in radio
or soft X-rays \citep{2008ApJ...686.1094H,2001A&A...365L.225B,2009PASJ...61.1217M}, where they are much less prominent.  The angular
resolution and sensitivity of \nustar\ were insufficient to allow
a detailed search for corresponding small-scale features at optical
wavelengths, where proper motions and/or radial velocities might give
clues as to the nature of these hotspots of electron acceleration.

\begin{figure}
\begin{center}
\includegraphics[width=1.0\linewidth]{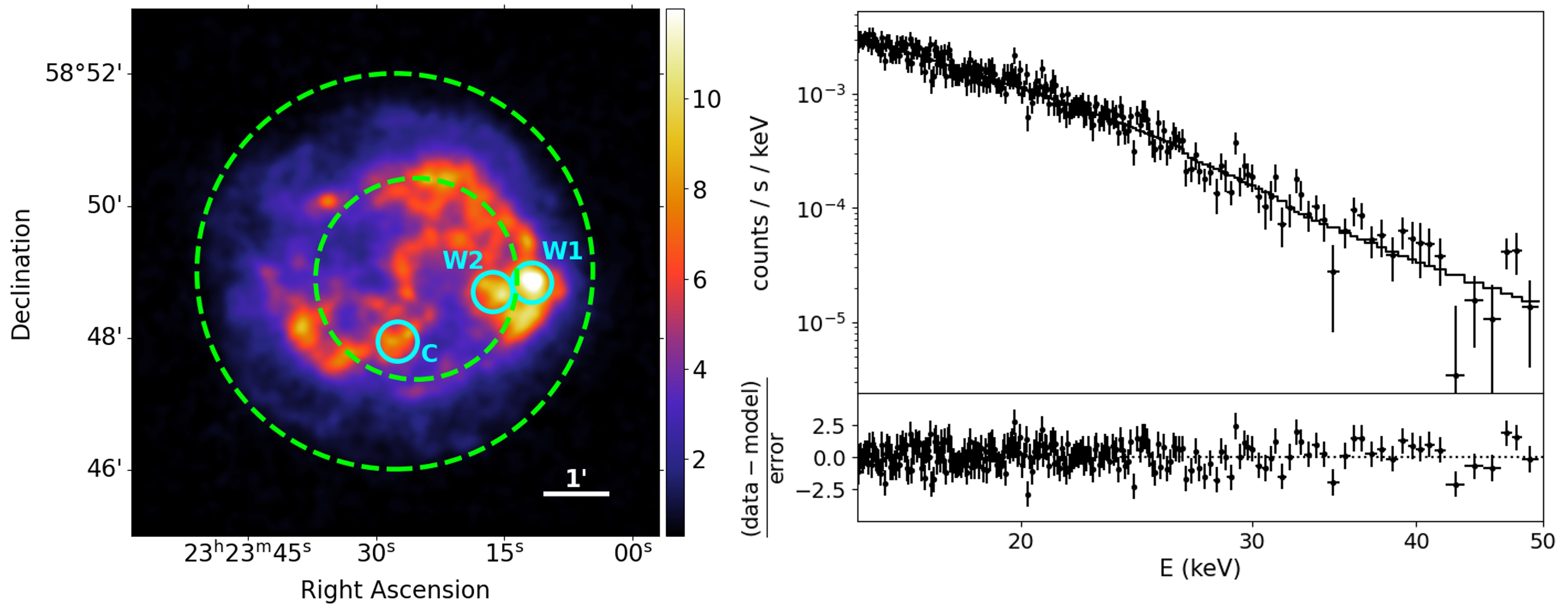}
\caption{\textit{Left}: Simulated HEX-P image of Cas A (15 -- 50 keV) for a 200 ks exposure. The green dotted lines mark the forward (outer ring) and reverse (inner ring) shock, respectively. The cyan circles (C, W1, W2) are the brightest hard X-ray knots detected by \textit{NuSTAR} \citep{NuSTAR_CasA15}. \textit{Right}: Simulated HEX-P spectrum of Cas A (W1 region in the left figure, $r=18''$) for 200 ks exposure. We input a power-law model with the best-fit parameters determined by archival \textit{NuSTAR} data ($\Gamma = 3.32$).  Fits to the simulated spectrum recovered the input value of the photon index with 2\% precision ($\Gamma = 3.38 \pm 0.06$).}
  \label{fig:casa}
\end{center}
\end{figure}

HEX-P, with improved angular resolution and sensitivity, will allow for resolving faint features such as shown in Figure \ref{fig:casa}. Non-thermal spectra will be mapped at the forward shock, reverse shock, and interior of the remnant. This spatially resolved spectroscopy will provide the best opportunity to study the variability on timescales of a few years of non-thermal emission at the hard X-ray knots suggested by \cite{Chandra_CasA} and \cite{2008ApJ...677L.105U}. Such flux variability requires enhanced magnetic fields ($0.1-1$ mG) perhaps at ``inward shocks'' induced by a density gradient due to molecular clouds \citep{CasA_inward} or asymmetric circumstellar material (\cite{2022A&A...666A...2O} and references therein). While \cite{Chandra_CasA} left open the possibility of variable thermal emission ($kT \sim 1-3$ keV) from \chandra\ observations, the high-energy sensitivity of HEX-P HET will minimize the complication with thermal emission and probe clean non-thermal continuum with photon index $\Gamma=3-3.5$ above 15 keV as seen by \textit{NuSTAR}. Independent estimates of the magnetic field \citep[such as those based on ``thin rim" morphology;][]{parizot06} are also of order 0.2 mG, giving synchrotron cooling timescales of 1 -- 10 years.  
Our simulation shows that a 200-ks observation with HEX-P can measure the flux of the hard X-ray knots ($r=18''$ circles) at 2\% precision. At this level of angular resolution and precision of flux measurement, the magnetic field can be accurately measured not only at the inward shock but also at other smaller hard X-ray knots throughout the entire remnant. Such a study will reveal the magnetic field structure of the remnant at the location of the  highest-energy electrons, in synergy with the polarization measurement by IXPE \citep{2022ApJ...938...40V}.
The connection between such observations and the highly detailed
radio polarimetric studies, which reflect the much larger regions populated by GeV-energy electrons, can help in following the evolution
of nonthermal particles beyond their acceleration sites and in filling
out our general picture of the production and evolution of relativistic
particles in a supernova remnant.



\subsection{Tycho}

SNR G120.1+1.4, famously known as Tycho, belongs to a class referred to as ``Historical Supernovae" due to its association with the supernova explosion SN 1572.  Its classification as the remnant of a normal Type Ia supernova was confirmed with light echos \citep{2008Natur.456..617K}. 
Tycho was one of the first Galactic radio sources identified (``Cassiopeia B"), shown to be separate from the much brighter nearby Cassiopeia A by \cite{1952Natur.170..364H}, and was only the second radio source, after the Crab, to be identified with a known supernova.  That radio synchrotron emission demonstrates the presence of electrons accelerated to GeV energies. Non-thermal X-ray emission was detected up to 25 keV with HEAO-1 \citep{1979ApJ...234L.195P}, indicating for the first time the presence of electrons with TeV energies in Tycho. Following this detection, \chandra\ observations located most 
non-thermal emission in narrow filaments around Tycho's rim \citep{2002ApJ...581.1101H,2005ApJ...621..793B,2005ApJ...634..376W}. Following that, several non-thermal ``stripes" inside Tycho were revealed by follow-up deeper \chandra\ observations \citep{2011ApJ...728L..28E}.  
The spacing of these stripes was argued to be of the order of the gyroradius of the highest-energy protons present, giving (for a magnetic field of order 30 $\mu$G) proton energies above $10^{15}$ eV, near the ``knee" of the cosmic-ray spectrum around 3 PeV. Subsequent variability
studies \citep{2020ApJ...894...50O} gave an even higher estimate
of $\sim 100\ \mu$G, but the relation to the maximum proton energy
relies on several untested assumptions.

\cite{2015ApJ...814..132L} used \nustar\ to perform a spatially resolved spectroscopic analysis of the synchrotron emission and radioactive $^{44}$Ti in Tycho's SNR using  a deep ($\sim$ 750 ks) \nustar\ observation. The hard ($>10$ keV) X-rays were found to be concentrated in the southwest of the remnant, where the earlier \chandra\  observation had found the high emissivity ``stripes".  No evidence was found for $^{44}$Ti, and only upper limits were put on its presence. Spatially resolved spectra
were fit with a simple model of synchrotron emission from a power-law
electron energy distribution with an exponential cutoff, which produces
a slower-than-exponential rolloff of emission characterized by the frequency $\nu_{\rm roll}$ at which the spectrum has dropped by a factor
of 10 below its extrapolation from lower frequencies.  Values of rolloff energy $h\nu_{\rm roll}$ were found to vary over a factor of 5 at different
regions around the rim \citep{2015ApJ...814..132L}, and were shown to rise as a steep function of the shock velocity obtained from radio expansion
measurements \citep{1997ApJ...491..816R}.

In addition to Tycho's well-characterized X-ray emission, gamma rays were detected by \fermi-LAT at GeV energies \citep{2012ApJ...744L...2G} and by VERITAS at TeV energies \citep{2011ApJ...730L..20A}. The detection in the TeV range proves that efficient particle acceleration to very high energies is taking place, presumably at the SNR's shock. Whether the gamma rays are
due to leptonic or hadronic processes is not clear.
 The gamma-ray spectrum of Tycho was  modeled and found to be consistent with diffusive shock acceleration (DSA) of electrons, in a two-zone leptonic model,  \citep{2012ApJ...749L..26A}, but a hadronic model invoking shock-accelerated cosmic-ray protons was also possible \citep{2013ApJ...763...14B}. HEX-P will allow better constraints
 on high-energy spectral structure which will aid in this discrimination.

\begin{figure}[ht!]
  \centering
  \includegraphics[width=0.5\linewidth]{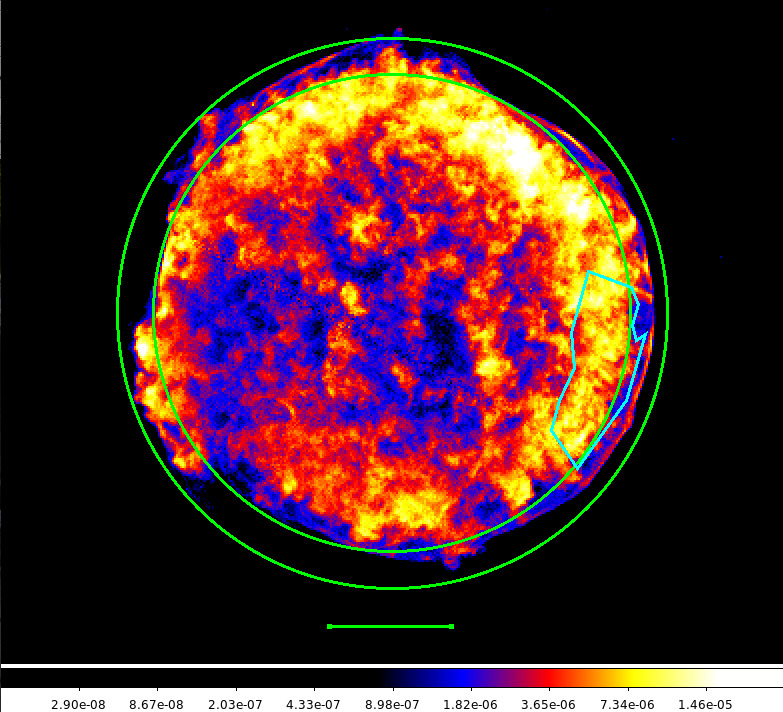}
  \caption{\chandra\ image of Tycho (0.7 -- 9 keV), binned
  by a factor of 4.  Brightness scale is logarithmic; units
  are erg cm$^{-2}$ s$^{-1}$.  The scale bar has length $2'$. The annulus
   defines the ``rim" region, which contains a high percentage of nonthermal emission.  The irregular cyan region in the west has
   a harder spectrum than most of the remnant.  Simulated HEX-P spectra
   from each, based on \chandra\ 
   data, are shown below.}
\label{fig:tychoregions}
\end{figure}

\begin{figure}[h!]
\begin{center}
\includegraphics[width=1.0\textwidth]{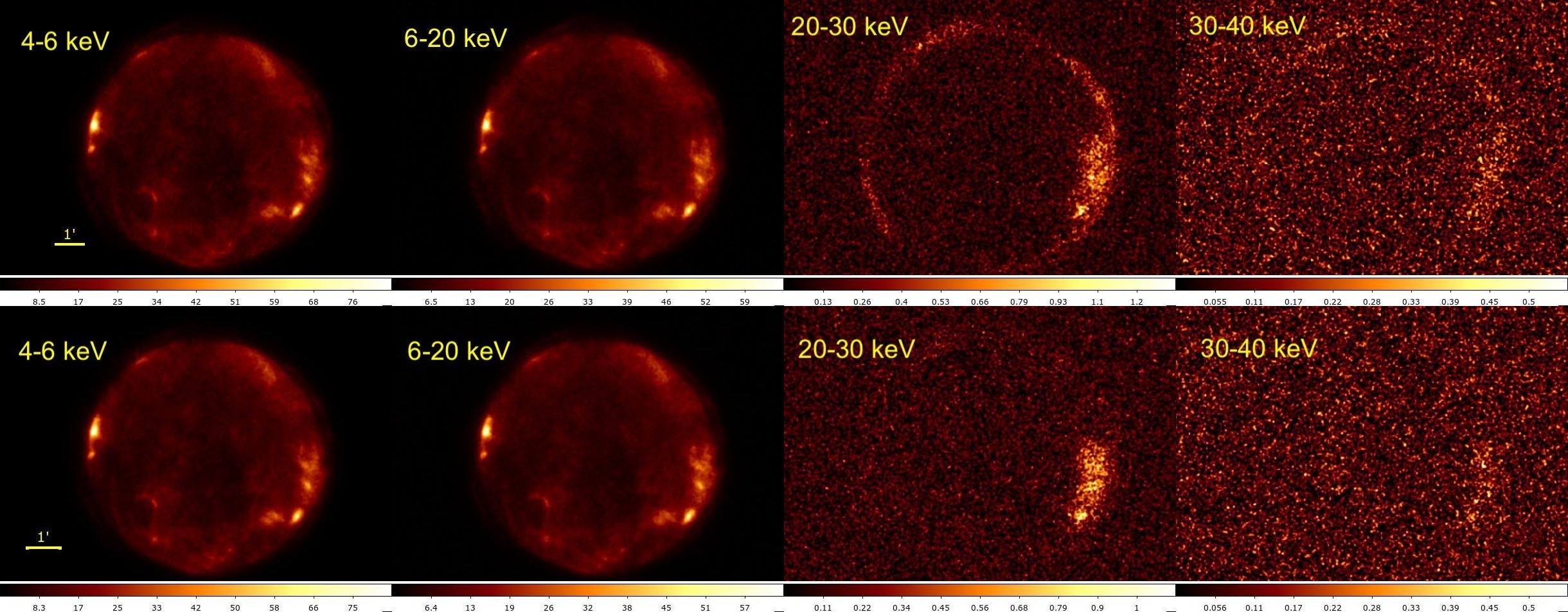}
\end{center}
\caption{Simulated 100 ks HEX-P images of Tycho's SNR. Input spectral models include a power-law for the W hard interior region, one of two power-law models for the rim region ($\Gamma = 2.67$ or 3.48) and a thermal shock model for the remaining interior (parameters
not specified as this component contributes negligibly above 8 keV).
Units are integrated counts in the indicated energy ranges.
Upper row: Rim power-law, $\Gamma = 2.67$. Bottom row: Rim power-law model, $\Gamma = 3.48$.   
The different rim models can clearly be distinguished.}
\label{fig:tycho_sixte_sim}

\end{figure}
  \subsubsection{Imaging and Spectral Simulations }

Figure~\ref{fig:tychoregions} shows the broadband (0.7 to 9 keV) \chandra\ image of Tycho.  Regions previously identified with strong nonthermal emission include the thin rims seen around most of the periphery, as well as structures in the west. \chandra\ spectra  from the annulus (``rim") and irregular hard-spectrum west region (``W hard") were extracted and fit with power-law models, which were
used to simulate images in the LET and HET instruments.  
Models for the rim were power-laws with photon index $\Gamma = 2.67$ and 3.48 (compared in Figure~\ref{fig:tycho_sixte_sim}).
The same model for the W hard region was used in each. A purely thermal spectrum of the remainder of the remnant was assumed.  The models can clearly be distinguished; the rim is clearly resolvable 
separately from the W interior region, allowing model discrimination not possible with the \nustar\ sensitivity and angular resolution.

\begin{figure}
\begin{center}
\includegraphics[width=1\linewidth]{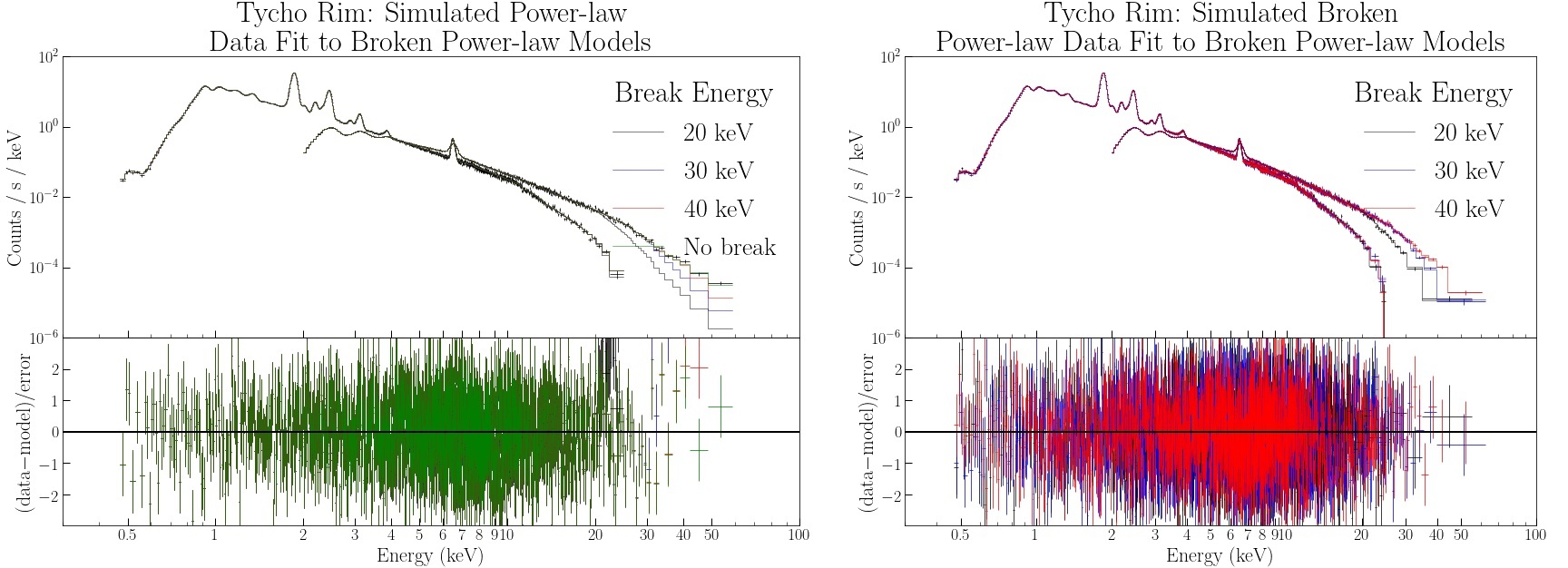}
  \caption{Simulated spectra for a 200-ks observation of Tycho.  Left: Simulated spectra from parameters given by pure power-law fits to \chandra\ rim data, 
  fit with one pure and three broken power-law models for a 200-ks observation of Tycho.  C-statistic values are 398.0 for the power-law fit, and rapidly deteriorate for fits using broken power-law models:   433.7, 662.4, and 2273.8 
  for increasingly severe break energies (40, 30, and 20 keV respectively), clearly ruling out the broken power-law models. Right: Simulated spectra from parameters given by broken power-law fits to the \chandra\ data, fit with broken power-law models.  Input values are recovered to $\approx 2.5\%, \approx 5\%$, and $ \approx 10\%$ accuracy for break energies of 20, 30, and 40 keV, respectively (Cstat values of 418.9, 414.2, and 414.0.}
  \label{fig:tychospec}
\end{center}
\end{figure}

Figure~\ref{fig:tychospec} shows simulated spectra using XSPEC and the V7 response files for the HET and LET.  In all cases, an input
power-law index of $\Gamma = 3.0$ was assumed, with normalization
based on the \chandra\ image.
All spectra are consistent with the data in the \chandra\ band, but with different assumptions about behavior at higher energies.
In a 200-ks exposure, parameters of XSPEC simulated straight power-laws can be recovered to within $\sim 5\%$ accuracy, while models
with broken power-laws (with $\Delta \Gamma = 3$) do a distinctly poorer job, as quantified by the increase in C statistics.  
A second set of input models, broken power-law simulated data also steepening by $\Delta \Gamma = 3$ at various assumed energies $E_{\rm break}$, produce the simulations shown in Figure~\ref{fig:tychospec}.  
Fits with the same models recover the two photon
indices and $E_{\rm break}$ to within $\approx 2.5\%, \approx 5\%$, and $\approx 10\%$ for break energies of 20, 30 and 40 keV, respectively. These accuracy levels will enable the detection of hard X-ray variabilities from several prominent X-ray features where particle acceleration is considered to be more energetic and active. These simulations demonstrate the ability of the HET, especially to characterize the spectral shape in hard X-rays from Tycho, with important implications for the nature of electron acceleration.  The detection of steepening, either as smooth curvature
or more abrupt spectral breaks, is essential for determining the
maximum energies of electrons.

\subsection{G1.9+0.3}

\begin{figure}[h!]
\begin{center}
\includegraphics[width=1.0\textwidth]{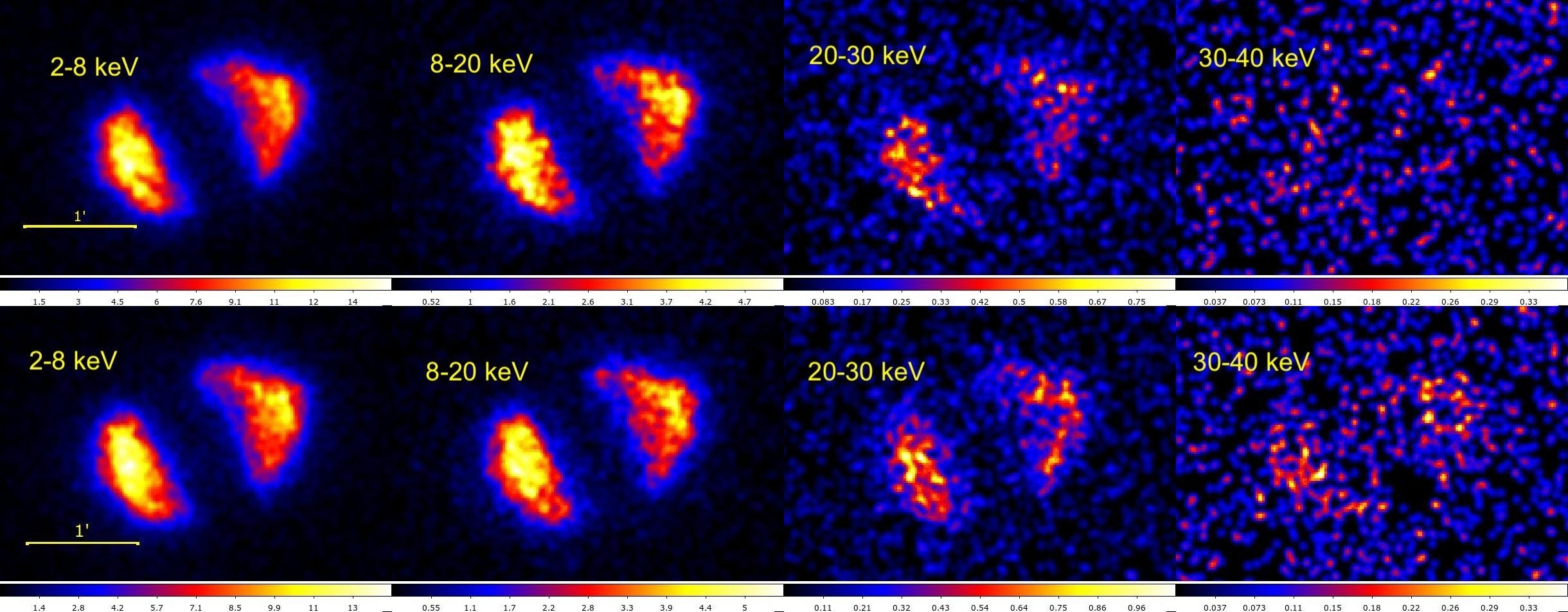}
\end{center}
\caption{Simulated 100 ks HEX-P images of G1.9+0.3, based on two spectral fits indistinguishable to \nustar\ \citep{zoglauer15}. Upper row: Power-laws with an exponential cutoff in photon energy (E limb: $\Gamma = 2.09$, $E_{\rm break} = 14.9$ keV; W limb, $\Gamma = 2.07$, $E_{\rm break} = 13.9$ keV. Bottom row: Straight power-laws (E limb: $\Gamma = 2.59$; W limb, 
$\Gamma = 2.66$).  Units are counts.}
\label{fig:g1p9}
\end{figure}

G1.9+0.3 is the remnant of the most recent supernova in the Milky Way
\citep{reynolds08}, with an expansion age of order 140 years (an upper
limit to the true age) and (based on hydrodynamic models of the
deceleration) a likely explosion date of around 1900
\citep{carlton11}.  It has an X-ray spectrum dominated by synchrotron
emission, with one of the highest rolloff energies of any SNR
\citep[over 2 keV;][]{reynolds09b}.  However, thermal emission is
detected in small regions, with line widths of up to 14,000 km
s$^{-1}$ \citep{borkowski10}, the fastest seen in any Galactic SNR,
and consistent with the measured expansion at an assumed distance of
8.5 kpc.  The remnant is sufficiently young that particle acceleration
(at least in conventional theories) is limited by the remnant age --
that is, the maximum electron energy is not set by synchrotron losses,
and is the same as the maximum ion energy. This would make G1.9+0.3 a
unique case where the maximum ion energy can be inferred directly from
X-ray observations.  The remnant is also the only Galactic SNR still
increasing in brightness, at both radio and X-ray wavelengths
\citep{carlton11}, and it is expected that the maximum particle energy
is also continuing to increase.  \nustar\ observations \citep{zoglauer15} showed the
presence of X-rays out to about 25 keV before the signal was lost in
the background.  Those observations were unable to discriminate
between straight power-laws and power-laws with relatively low ($\sim
15$ keV) cutoff energies, a crucial difference in characterizing the
highest-energy electron spectrum.  As the simulations of Fig.~\ref{fig:g1p9} show, HEX-P can
resolve this problem.  As time goes on, G1.9+0.3 provides us with a
view of the evolving process of particle acceleration.

\subsection{SN1987A} 

SN1987A is the closest supernova known since its explosion 36 years ago in the Large Magellanic Cloud (LMC) located 51 kpc away. While it has been best known for the detection of neutrinos from the supernova event, SN1987A has exhibited  spatial and spectral evolution across a wide range of wavelength bands from radio to X-ray. Shortly after the explosion, X-rays were detected from SN1987A (e.g., \citet{Frank2016}). The X-ray emission, composed of soft and hard components which vary separately, is primarily thermal, considered to arise from the interaction between the expanding shock wave and the surrounding circumstellar material. 

\begin{figure}[ht!]
    \centering
\includegraphics[width=0.6\linewidth]{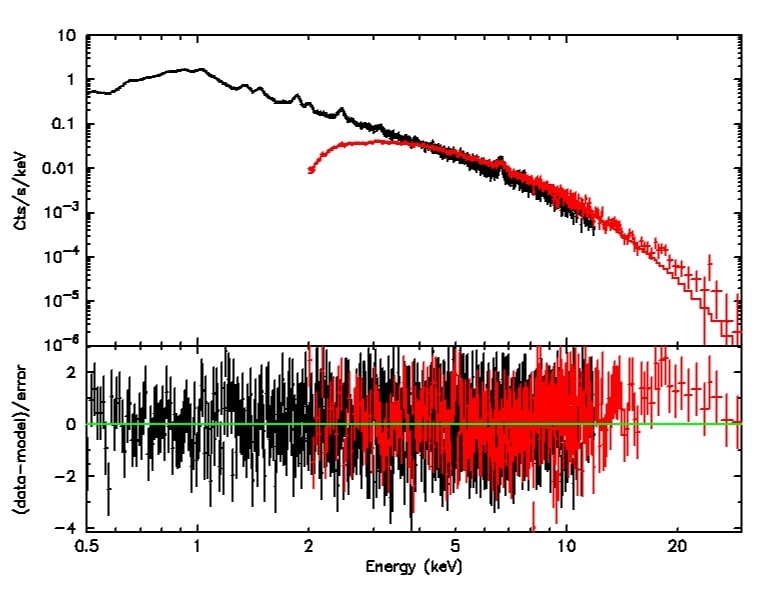}
    \caption{Simulated HEX-P LET (black) and HET (red) spectra of SN1987A assuming a 300 ks exposure. The input spectral parameters for the three thermal components ($kT_1 = 0.48, kT_2 = 0.96, kT_3 = 3.2$ keV) and a power-law component ($\Gamma = 2.6$) were adopted from the \chandra, \xmm\ and \nustar\ observations obtained in 2020 \citep{Greco2022}. The simulated \hexp\ spectra are fit with the thermal components only to illustrate the presence of the non-thermal emission above 10 keV. The non-thermal component is detected with $7\sigma$ significance and its photon index is well constrained to $\Gamma = 2.4_{-0.1}^{+0.2}$, much improved from the 2020 \nustar\ observation with 350 ks exposure ($\Gamma = 2.6_{-1.8}^{+0.7}$) \citep{Greco2022}.     
    }
    \label{fig:sn1987_spectra}
\end{figure}

One of the key questions associated with SN1987A is whether the supernova left a neutron star or black hole. Some hints on the presence of a NS at the core were suggested by recent ALMA  radio and \nustar\ hard X-ray observations. ALMA detected a bright spot, suggesting that local dust is being heated by the NS's thermal emission \citep{Cigan2019}. \nustar\ observed non-thermal X-ray emission above 10 keV, suggesting the presence of a PWN \citep{Greco2021}. However, despite  the \nustar\ observations with 3.3 Ms total exposure, the detection of a PWN component is still under debate \citep{Alp2021, Greco2022}, and more sensitive hard X-ray observations are required for making a firm detection of the putative PWN.  A contribution to the
hard X-rays from synchrotron emission from electrons accelerated in the SN blast wave cannot be ruled out, as the blast wave radius is smaller than an arcsecond. Note that only the hard X-ray band above 10 keV can be observed from the core region due to severe photo-absorption by heavy-element ejecta \citep{Alp2018}. HEX-P's energy
coverage from 0.2 to 80 keV will be essential for fully characterizing both thermal and non-thermal X-ray components. Figure \ref{fig:sn1987_spectra} shows simulated HEX-P spectra of SN1987A. While SN1987A is a point source to HEX-P as it is for \nustar, HEX-P's better angular resolution means smaller background in a resolution element, enhancing the sensitivity to unconfirmed non-thermal emission above 10 keV compared to \nustar.

Since the multi-wavelength emission components from SN1987A are expected to vary on year timescales, as seen over the last 35 years, HEX-P will be able to provide a broadband and high-resolution X-ray view of how both thermal and non-thermal X-ray emission appears and evolves in the 2030s. Since CTA selected the LMC survey as one of CTA's key science projects \citep{CTA2023},  HEX-P and CTA will jointly reveal hard X-ray and TeV emission from synchrotron and 
inverse-Compton radiation associated with the SNR and possibly emerging PWN,  respectively.   

\section{Pulsar wind nebulae} 
A pulsar-wind nebula (PWN) is a bubble of non-thermal radiation from a magnetized, primarily electron/positron plasma which is supplied by an energetic pulsar. The confinement of the initially relativistic pulsar wind by its environment results in a termination shock. Particles are thought to be accelerated to high energies in the shock and flow and diffuse outward to produce a PWN. PWNe are expected to evolve through three phases: an initial
phase in which the young PWN expands into the cooled expanding ejecta of its natal SNR; a second phase in which the SNR's reverse shock wave has returned toward the center and recompressed
the PWN, distorting it and perhaps moving it off-center in the SNR; and a third phase,
after the pulsar has either left the SNR or outlived it, and its wind is interacting
with undisturbed ISM.  Since pulsars are typically born with substantial kick velocities,
in this third phase they are normally moving supersonically in the ISM and their
PWN is in the form of a bow-shock nebula.  Most modeling has been done for the
earlier stages.  \citet{kennel1984a,KC1984}  
modeled the Crab Nebula with a spherically symmetric magnetohydrodynamic model assuming pure advection of particles behind the termination shock, successfully explaining its basic features.  Later detailed MHD simulations have reproduced many features of the X-ray morphology of the Crab Nebula, including small-scale structures such as jets and tori \citep[e.g.,][]{komissarov2004,delzanna2004}, features found in many other PWNe \citep[][and references therein]{ng04}.
While these MHD models capture the main properties of the Crab Nebula, they do less well in describing properties common in other PWNe that the Crab does not share, such as uniformly steepening spectra with distance from the pulsar.  In general, there is a great deal more to be learned about the diversity of PWN phenomenology that the MHD models cannot answer. TeV and even PeV photons are observed in PWNe \citep[][]{Cao2021}, meaning that they accelerate electrons to even higher energies. These particles can escape from the PWNe and could contribute significantly to the flux of energetic electrons and positrons detected on Earth \citep[e.g.,][]{abeysekara2017}.

Particle acceleration in PWNe is thought to occur at the relativistic termination shock via DSA or magnetic reconnection as demonstrated by particle-in-cell (PIC) simulations \citep[e.g.,][]{sironi2009,Sironi2014}. However, it is still unclear to what energy the particles are accelerated, how they evolve as they flow and diffuse, and how they escape from the PWNe. These can be addressed by high-quality X-ray and gamma-ray data. In particular, synchrotron X-rays probe the highest-energy electrons, as their inverse-Compton gamma-ray emission is suppressed by
the Klein-Nishina effect. Sensitive X-ray and gamma-ray observatories have helped to address these questions during the previous two decades, but many important questions remain unanswered. We list questions (1)--(3) as shown below.

\subsection{Important questions to address on PWN physics} 

1. Particle acceleration in relativistic shocks \\
Numerous PWNe emit TeV (even PeV) photons, implying that there must be very energetic particles accelerated at their termination shocks or elsewhere.
Indeed, half of the unidentified TeV sources discovered in H.E.S.S. Galactic plane surveys are now categorized as middle-aged or old PWNe \citep{hess18}, and half of the sources listed in the first LHASSO source catalog are linked with pulsars and/or PWNe \citep{lhasso23}. Most of them are theoretically compatible with being powered by a pulsar \citep[][]{Wilhelmi2022}.
Because the energy distribution of the particles is imprinted in their radiation spectrum, accurate measurements of PWN spectra will give us clues to particle acceleration processes in relativistic shocks. Although PeV-energy particles are already detected in the Crab Nebula, an important question is whether or not other PWNe can generally accelerate particles to such energies (i.e., constitute leptonic PeVatrons).  As in other sources, Klein-Nishina suppression of inverse-Compton scattering from the highest-energy electrons means that their energy distribution, up to those highest energies, is best probed not by gamma rays but by sensitive high-energy X-ray observations of synchrotron emission.
Modeling of X-ray and gamma-ray data of various PWNe has given maximum particle energies of $\sim$PeV not only in young sources \citep[$\simlt 10^4$\,yrs; e.g.,][]{Torres2014, Abdelmaguid2023} but also in middle-aged ones \citep[$\sim 10^4$--$10^5$\,yrs; e.g.,][]{Burgess2022,Park2023a,Woo2023}. While these estimates depend strongly on $B$ within the PWN (an electron of energy $E$ emits its peak synchrotron power at a photon energy
$h\nu_m \propto E^2\,B$), broadband spectral-energy-distribution (SED) models can constrain $B$, especially if spectral features such as spectral breaks or cut-offs, spatial variation of the spectrum, and/or energy-dependent morphology are measured. An independent $B$ estimation is possible if the seed-photon radiation energy density $u_*$ in the PWN region is known, since $P_{\rm synch}/P_{\rm IC}=u_B/u_*$ (in the Thomson regime), with $u_B \equiv B^2/8\pi$ is the
energy density in magnetic field.  The state-of-the-art code {\tt GALPROP} \citep[][and references therein]{Porter2022} provides models for Galactic distributions of cosmic rays as well as of the interstellar radiation field.\footnote{https://galprop.stanford.edu/code.php?option=theory} The latter provides models for $u_*$ within the Galaxy, thereby allowing an independent estimation of $B$ in PWNe. High-quality X-ray data will facilitate measurements of the aforementioned spectral and spatial features with high precision, and these, combined with improved $u_*$ estimations, will improve our understanding of particle acceleration processes in PWNe.

2. Particle injection into the interstellar medium \\
The origin of the cosmic-ray positron flux detected on Earth at energies higher than a few GeV is a longstanding puzzle \citep[e.g.,][]{Aguilar2019}, and it was suggested that nearby pulsars and their PWNe are an important source of these particles \citep[e.g.,][]{Cholis2013,DellaTorre2015,Xi2019,Manconi2020}. Radiative energy losses restrict possible sources of the highest-energy terrestrial cosmic-ray leptons with $E \simgt 1$ TeV to within about 1\,kpc, but studying numerous PWNe in a larger volume can aid in understanding energetic leptons within the Galaxy. After being accelerated at the relativistic termination shock of the pulsar wind, leptons evolve in the PWN under radiative and advective cooling \citep[e.g.,][]{Reynolds09,Van_Etten_2011,Park2023b}.
Young systems are expected to efficiently confine the nebular plasma, and in general they might not be an efficient source of energetic cosmic-ray electrons in the Galaxy, since their electrons rapidly lose energy to radiation via synchrotron emission in the strong magnetic fields typical of such objects.
However, we do have direct evidence of an efficient escape of particles in the outer medium from middle-aged and more evolved systems \citep[see e.g. ][for a recent review of systems in different phases and their observational signatures]{Olmi2023}.

A first clear indication is TeV halos seen around old PWNe \citep[e.g.,][]{Yuksel2009,bamba10,sudoh19,Abeysekara2017b,HESS2023}. These TeV halos are likely due to the sum of all particles which have escaped over time during the lives of the PWNe.  
A second piece of evidence  is the presence of extended (from 1 to several pc), asymmetric and collimated X-ray features protruding from some high-speed bow shock PWNe \citep[see, e.g.,][]{Kargaltsev2017}, showing a strong misalignment with respect to the pulsar direction of motion. These intriguing features can be explained as formed by energetic particles leaving the PWN, advected away along the ambient magnetic field lines \citep[][]{Olmi2019}.
Some degree of amplification of the ambient field is necessary to explain the observed emission as due to synchrotron radiation \citep[][]{Bandiera2008}, possibly produced by the particles themselves.
An excellent example is the Lighthouse PWN (IGR~J11014$-$6103) powered by the old 120\,kyr pulsar PSR~J1101$-$6101 \citep[][]{Halpern2014}. It shows a $\sim 6’$ long and bright collimated outflow to the west \citep[][]{Pavan2016}, with evidence for spectral evolution along the feature in the combined observations from \chandra\ and \nustar\ \citep[][]{Klinger2022}.
The possibility of investigating the spectrum of such features in the very extended energy band of HEX-P, and with its angular resolution,  is a key ingredient to understand not only the properties of escaped particles (and thus of the PWN) but also how they interact with the ambient medium, with important consequences for our understanding of the physics of transport and diffusion in the vicinity of sources.
All these points are then intimately connected both with point (1) —  how and to what energy particles are accelerated at the pulsar wind termination shock — and the following point (3) — how particle acceleration and escape change as the PWN evolves.

3. Co-evolution of the PWN and its SNR \\
The evolution of a PWN is tightly linked to that of its host SNR. The entire evolutionary history affects the particle energy distribution within the PWN today, which is imprinted in the emission SED. The size and expansion speed of PWNe provide additional information on the co-evolution since the PWN expansion is influenced by, first, the surrounding SNR ejecta, and later by the reverse shock \citep[e.g.,][]{Hattori2020}. Effects of PWN/SNR co-evolution \citep[e.g.,][]{Bandiera2020, Bandiera2023} are particularly important in understanding cosmic-ray electrons since we need to consider the entire history of particle injection by PWNe to compare with cosmic rays detected today. The co-evolution is very complex, and hydrodynamic (HD) simulations have been carried out to study the dynamic evolution of PWN-SNR systems \citep[e.g.,][]{Bucciantini2003,Kolb2017}. In these works, the broadband spectral properties of the PWNe have not been considered. \citet{Gelfand2009} developed a dynamic and radiative evolution model by treating the PWN-SNR co-evolution in a semi-analytic way \citep[][]{Reynolds1984}, in a one-zone model with spherical symmetry. This evolutionary PWN model has successfully fit the SED, size, and expansion speed of some PWNe \citep[e.g.,][]{Abdelmaguid2023}, and the model-inferred properties of the SN and SNR seem to agree reasonably well with other SNR evolution models (e.g., HD simulations). While PWN evolution models are very useful for understanding PWN physics and can provide a supplementary tool for understanding SNR physics, the models have many covarying parameters, and one-zone models fail to capture the obvious inhomogeneities and morphological complexities of almost all PWNe.
Moreover, recent higher-resolution HD studies \citep[e.g.,][]{Bandiera2020, Bandiera2023} found deviations from the prescriptions used for the dynamic evolution in the radiative evolution model of \citet{Gelfand2009}. These deviations may be manifest in the spatially-varying emission properties and can have a significant impact on the highest-energy particles. Measurements of the spatially-integrated SED, size, and expansion speed are certainly insufficient for studying the co-evolution in detail and precise measurements of the PWN properties (e.g., spatially-resolved spectra) are necessary to enrich our understanding of PWNe and SNRs.

\subsection{HEX-P contribution to PWN physics}
Accurate characterizations of the PWN emission properties are crucial to address the aforementioned questions. Question 1 requires sensitive measurements of high-energy X-ray spectra of PWNe. For Questions 2 and 3, accurate measurements of spatially-resolved emission properties are necessary, along with 
an emission model that takes into account the temporal and spatial evolution of particles within the PWN. While existing X-ray observatories have helped address these questions, their sensitivity to faint emission and to spectral detail is not yet sufficient to provide firm answers. \chandra\ and \xmm\ have measured PWN images and spectra in great detail, but their lack of sensitivity above 10 keV precludes detailed investigation of the most energetic electrons. While this could be supplemented by \nustar's hard X-ray data \citep[see][]{Reynolds2016,Mori2021}, cross-calibration uncertainties between the observatories cause some problem \citep[e.g.,][]{Madsen2015,Madsen2017,Hattori2020,Abdelmaguid2023}. In addition, the low angular resolution (60$''$ HPD), limited effective area, and strong and inhomogeneous backgrounds of \nustar\ \citep[e.g.,][]{Wik2014,Madsen2017b} have been a concern for measurements of spatially-integrated and resolved spectra of PWNe above 20 keV.

\begin{figure}[ht!]
\begin{center}
\includegraphics[width=1.0\linewidth]{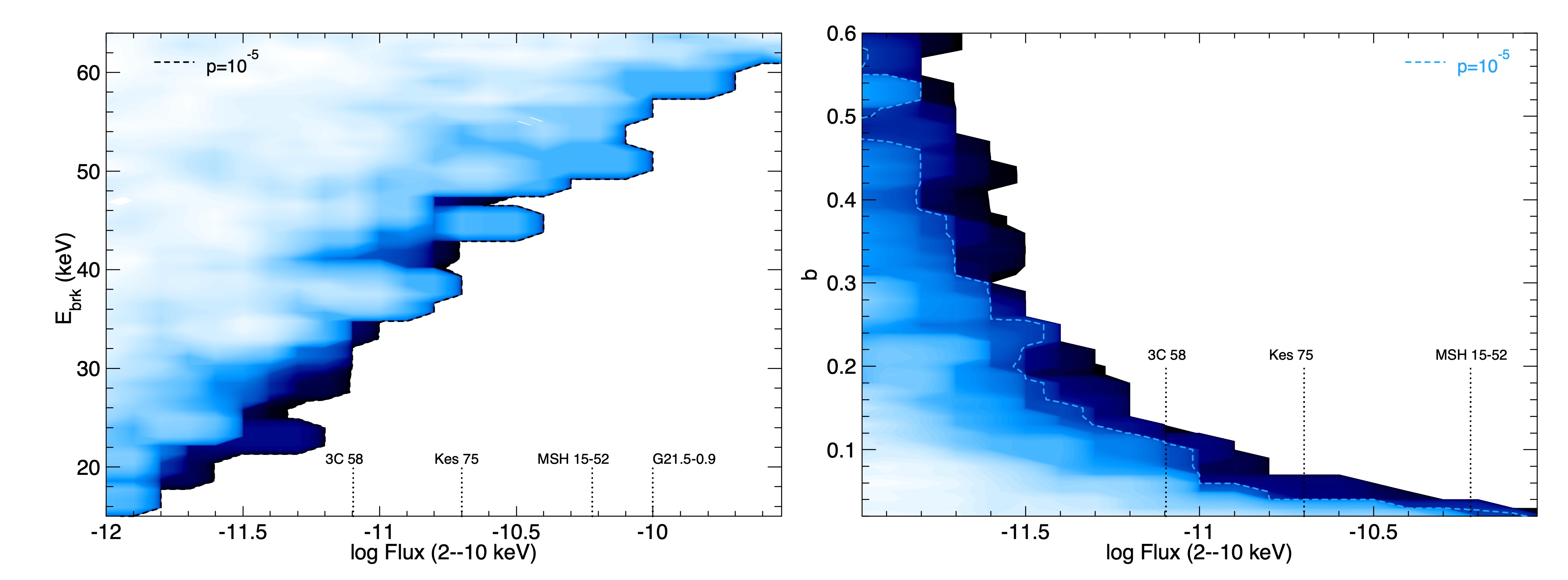}
    \caption{Results of HEX-P simulations of a generic $R=3'$ extended PWN. Left: $F$-test probability contours between a power law (PL) and a broken power-law (BPL) model. Simulated HEX-P data (100-ks) for a BPL spectrum having $\Gamma_l=2$ and $\Gamma_h=5$ at break energy $E_{\rm brk}$ are fit by a BPL and a PL, and the addition of the break was tested using the $F$ statistic. Right: Same as the left, but the simulations were carried out for a {\tt logpar} model with varying $b$ (see text). The vertical dashed lines mark fluxes of some young and bright PWNe: 3C~58 \citep[][]{An2019}, Kes~75 \citep[][]{Gotthelf2021}, MSH~15$-$5{\sl 2} \citep[][]{An2014}, and G21.5$-$0.9 \citep[][]{Nynka2014}.}
    \label{fig:R3mpwn}
    \end{center}
\end{figure}

The HEX-P observatory can revolutionize our studies of PWNe. The broadband coverage of the combined LET and HET instruments provides a large lever arm for spectral measurements, and thus the properties of emission by highest-energy electrons, which may be manifested as a spectral cutoff, can be well measured. Figure~\ref{fig:R3mpwn} shows detectability of a strong spectral break at $E_{\rm brk}$ \citep[broken from $\Gamma_l=2$ to $\Gamma_h=5$ as was suggested for 3C~58;][]{An2019}, simulated for a generic $R=3'$ PWN assuming a 100\,ks observation with HEX-P. Note that other gradual cut-off shapes, e.g., exponential cutoffs at $E_{\rm cut}$, are more easily detected because the spectral shape starts to change at low energies as compared to the broken power-law case. We also mark fluxes of some bright PWNe in the figure for reference. The simulations show that we can detect a spectral break up to 50\,keV if the source is bright (e.g., $F\ge 10^{-10}$\,\fluxcgs\ in the 2--10\,keV band) and if its spectrum indeed exhibits a spectral cutoff. This experiment will be particularly useful for some PWNe in which a spectral curvature or cutoff was measured \citep[e.g., Crab, G21.5$-$0.9, PSR~J1400$-$6325, and PSR~J1813$-$1749;][]{Madsen2015b,Nynka2014,hitomi18,renaud10,bamba22,Hattori2020} or suggested \citep[e.g., 3C~58 and N157B;][]{An2019,bamba22}. This will help address Question 1. Note that \nustar\ data have some systematic uncertainty due to the inhomogeneous background which is very difficult to assess for diffuse PWNe. HEX-P, in addition to having twice the effective area and lower backgrounds, would not suffer from such inhomogeneous background. Hence the improvement can be substantial.

In studying PWN emission using an evolutionary model, it is crucial to measure the emission SED as precisely as possible. This can provide additional information to break parameter covariance of the evolutionary model since the SED shape changes depending sensitively on the cooling history of the particles \citep[e.g.,][]{Gelfand2009}. In this regard, the broadband coverage of LET+HET of HEX-P is particularly helpful. The right panel of Figure~\ref{fig:R3mpwn} shows HEX-P's sensitivity to a small degree of curvature in the emission spectrum investigated using the {\tt logpar} model in {\it XSPEC} ($dN/dE=K(E/E_p)^{-a-b \mathrm{log}(E/E_p)}$ with $E_p=5$\,keV) for $R=3'$ PWN emission. Simulations of a 100-ks exposure show that HEX-P will be able to measure a shallow curvature with $b<0.1$ (corresponding to $\sim$20\% SED change over the 0.3--80\,keV band) if the source flux in the 2--10\,keV band is $\ge 10^{-11}$\,\fluxcgs. This is sufficient for measuring spectral breaks detected in some young PWNe, which will provide important clues to the particle acceleration mechanisms and PWN-SNR co-evolution. In particular, \citet[][]{Klinger2022} found in outflow regions of the Lighthouse PWN that the soft X-ray (\chandra) spectra are discrepant with the hard X-ray (\nustar) ones, implying that the effects of particle evolution (e.g., cooling) are best seen at $>$8\,keV. However, this could not be confidently stated because of cross-calibration issues and non-uniform background in the \nustar\ data. HEX-P observations will help to clarify this since the suggested spectral curvature is large enough for detection, and can shed light on the properties of the particles escaping the PWN and their effect on the ambient medium (Question 2).
\begin{figure}[ht!]
\begin{center}
\includegraphics[width=1.0\linewidth]{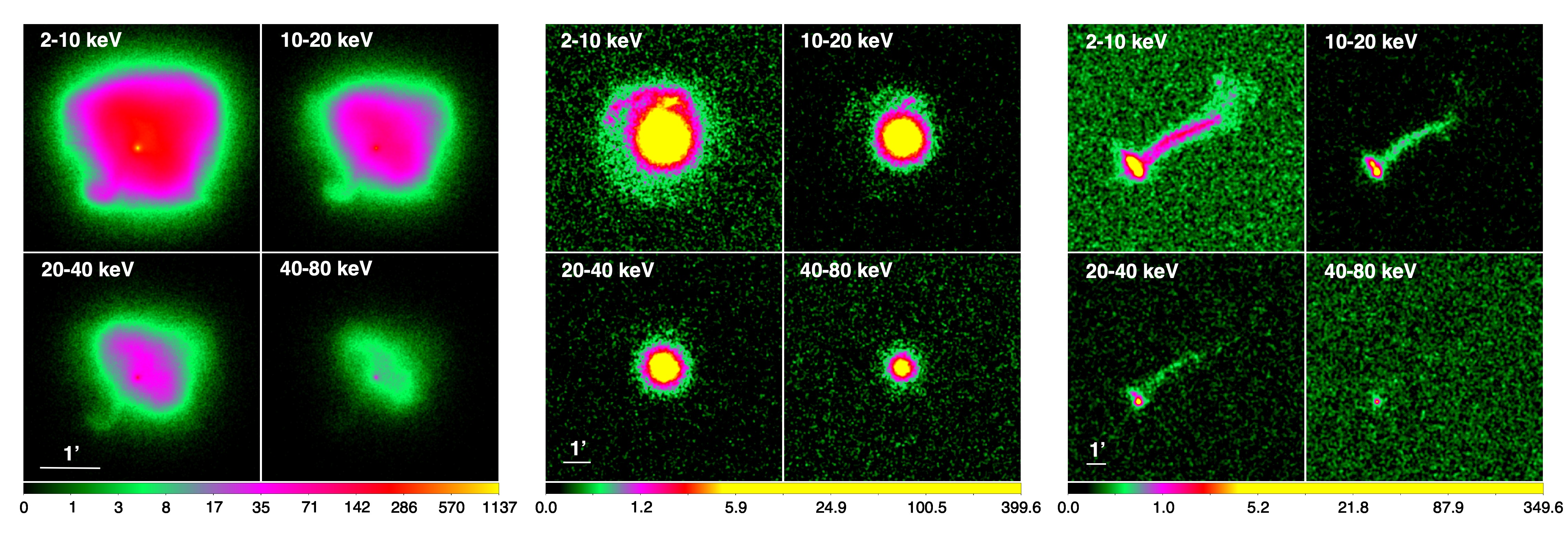}
    \caption{Simulated HEX-P HET images of the Crab Nebula (1\,ks; left), G21.5$-$0.9 (50\,ks; middle) and the Lighthouse PWN (right; 100\,ks) made with the SIXTE suite. Units are total counts; scales are logarithmic.  Chandra images were used as the inputs, and spectral variations within the PWNe as measured by Chandra \citep[][]{Mori2004,Guest2019,Pavan2016} were included.}
    \label{fig:Crabimage}
    \end{center}
\end{figure}

HEX-P will also allow measurements above 10 keV of faint emission in the outer regions of extended (and evolved) PWNe. This will help to characterize synchrotron burn-off effects as well as particle flow near the outer boundary of the PWNe, and to provide new insights into the evolution of PWNe after their young phase (Question 3). We can then estimate the amount and spectrum of electrons that escape from the X-ray PWN and propagate to larger TeV emission regions. With the advent of high-resolution TeV observatories (e.g., CTA), the HEX-P data will tell us about how the electron spectrum evolves as electrons propagate from the X-ray PWN to the TeV emission region, and eventually to the interstellar medium. This will help address Question 2. In our 100-ks simulations of a HEX-P observation of a $R=4'$ PWN, we found that we could measure the photon index of an annular region with the inner and outer radii of 3$'$ and 4$'$ to within 0.1 if the 2--10\,keV flux is higher than $3\times 10^{-13}$\,\fluxcgs. 
In other PWNe, the injection regions may be too faint for accurate measurement of the spectrum. These faint regions can still be identified by images, and our simulations of HEX-P observations for a hypothetical PWN suggest that faint source emission within an annular region with the inner and outer radii of 4$'$ and 4.25$'$ (i.e., width of 15$''$) can be detected with 3$\sigma$ confidence if the flux is higher than $3\times 10^{-14}$\,\fluxcgs.

\subsection{HEX-P simulations for the Crab, G21.5$-$0.9, Lighthouse PWN and G0.9+0.1}

HEX-P plans to observe the Crab, G21.5$-$0.9, Lighthouse PWN and G0.9+0.1 in the primary science program. The improved PSF of HEX-P HET will resolve substructures in the sources much better than \nustar\ could. In addition, HEX-P's large effective area will allow more accurate spectral characterization of the sources. Figure~\ref{fig:Crabimage} shows simulated images of the sources made with the SIXTE suite. Compared to \nustar's images of the sources \citep[][]{Madsen2015b,Nynka2014,Klinger2022}, the morphologies of the PWNe are much better resolved. The torus and jets of the Crab Nebula and the northern spur and eastern limb of G21.5$-$0.9 were seen by \nustar\ only after deconvolving the PSF from the images, and thus their spectra could not be measured. The HET of HEX-P can resolve those structures as in the simulated images (Figure~\ref{fig:Crabimage}), allowing spectral measurements.  Nonthermal emission from the northern spur and the eastern limb of G21.5$-$0.9 that were suggested based on deconvolved images \citep[][]{Nynka2014} can be better characterized by HEX-P observations. This also applies to other sources in which NuSTAR's resolution was insufficient to resolve various structures in them, requiring deconvolution \citep[e.g., G11.2$-$0.3;][]{Madsen2020}.

The Lighthouse PWN is particularly intriguing thanks to its long ($\sim 6'$) and misaligned tail, which connects to a diffuse emission region \citep[][]{Pavan2016}. The morphological change from a collimated jet to a diffuse region in this PWN can tell us much about electron injection from the old PWN into the ISM.
The \nustar\ measurement of the spectrum of the diffuse region ($\Gamma=2.21\pm0.08$) was discrepant with the \chandra\ result ($\Gamma=1.74\pm 0.05$) \citep[][]{Klinger2022}, suggesting that the electrons had experienced strong cooling. However, this was uncertain because of unknown systematic effects in the data, e.g., cross-calibration between the observatories and inhomogeneous background in the \nustar\ data \citep[the latter was a particular concern; see][]{Klinger2022}. 
Our HEX-P simulations for a 100-ks observation of the Lighthouse PWN, based on the previous \nustar\ results, found that HEX-P would be able to detect the diffuse emission at $\ge6'$ up to $\sim 20$\,keV (see Figure~\ref{fig:Crabimage} right) and to measure its spectrum accurately, e.g., constraining $\Gamma_X$ to within 0.04. This will provide sufficient sensitivity for discriminating between the hard {\it vs} soft injection spectra as measured by \chandra\ and \nustar.
More importantly, HEX-P data (LET+HET) will not suffer from systematic effects due to cross-calibration across the 0.3--80\,keV band or to inhomogeneous background. Hence, HEX-P observations of the Lighthouse PWN will considerably improve the \chandra\ and \nustar\ measurements, and thereby provide crucial clues to how the PWN electrons are injected into the ISM.

\begin{figure}[h!]
\begin{center}
\includegraphics[width=0.96\textwidth]{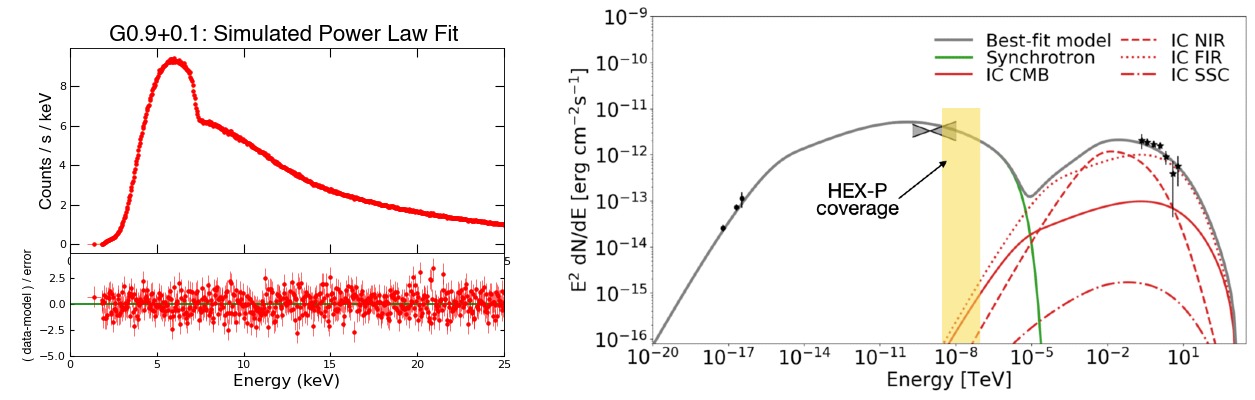}
\end{center}
\caption{\textbf{Left:} Simulated 50~ks HEX-P HET spectra of the absorbed non-thermal emission from the PWN G0.9+0.1. The spectrum is well detected up to 40 keV and all the parameters are recovered with $1\%$ or better.  
   \textbf{Right:} A one-zone time-independent SED model for G0.9+0.1 incorporating broadband data from existing radio, X-ray, and $\gamma$-ray data \citep{Fiori2020}.  The HEX-P energy band is shown in yellow, covering a critical portion of the SED. 
}\label{fig:g09_spec_sed}
\end{figure}

G0.9+0.1 is a composite-type supernova remnant (SNR) and pulsar wind nebula (PWN) located only one degree from the Galactic Center, long studied in the X-ray regime.   As the system is quite young with a spin-down age of $\tau_c=5.3$~kyr \citep{camilo09}, the SNR shell has not interacted with the $d\sim2'$ PWN core \citep{Helfand1987} leaving the nebula an ideal candidate for probing the physical mechanisms responsible for accelerating high-energy particles. X-ray observations below 10 keV by \chandra\ \citep{Gaensler2001} and \xmm\ \citep{Porquet2003, Holler12} reveal prototypical non-thermal emission with a radially dependent power-law index, a signature of PWN synchrotron cooling. Despite its young age and energetic pulsar, the nebula has coincident $\gamma$-ray emission 
\citep{Aharonian2005, Adams2021a} that is relatively underluminous for its pulsar luminosity, making this an interesting target to probe high-energy particle acceleration. \nustar observations were able to detect the non-thermal PWN emission up to 30~keV. However, stray light caused by the proximity to the Galactic Center reduces the quality of the data and the spectral parameters are only moderately constrained. HEX-P simulations of G0.9+0.1 show a significant improvement in the $>10$~keV range (Figure~\ref{fig:g09_spec_sed}, left). Broadband studies with multi-wavelength SEDs have attempted to probe the underlying acceleration processes.  Several one-zone time-dependent leptonic models \citep[e.g.,][]{Fiori2020} have been adopted with various constraints applied to input parameters, while other models explored lepto-hadronic scenarios \citep{Holler12}, multiple zones, or incorporated significant adiabatic cooling \citep{Tanaka2011}.  While these complex models fit the available data,  they often leave crucial parameters, such as the injected electron spectral index, unconstrained.  HEX-P provides critical coverage at the energy band where the majority of models predict a turnover in the SED curve (Figure \ref{fig:g09_spec_sed}, right).  HEX-P data added to broadband SED analysis will provide a deeper understanding of the extreme particle acceleration processes within PWNe. 

\subsection{Summary}
The broadband coverage, superior angular resolution, and low background of HEX-P HET will provide significant improvement in our understanding of PWNe. In addition to the spectral improvement, the HET of HEX-P will measure the morphologies of PWNe in greater detail compared to \nustar\ as presented in Figure~\ref{fig:Crabimage}. Hard X-ray emission from substructures (e.g., torus, jets, and knots) in PWNe will be better probed. HEX-P will observe the Crab Nebula and G21.5$-$0.9 as calibration targets, and these observations will provide unprecedented data for these two young and bright PWNe, allowing precise SED studies. In addition, a 100\,ks HEX-P observation of the Lighthouse PWN will measure the spectra of the outflow regions out to $\ge5'$ without systematic effects due to inhomogeneous background. This will probe how the electrons in the outflows evolve and merge into the ISM.


\section{Nucleosynthesis in the Galaxy}
%

\begin{figure}[h!]
\begin{center}
    \includegraphics[width=0.7 \textwidth]{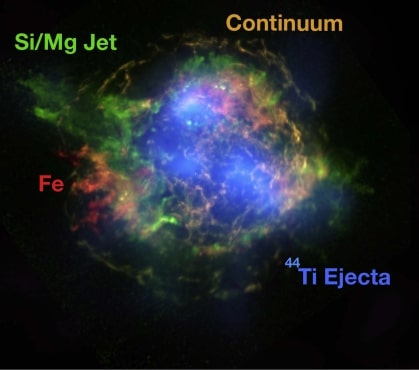}
    \end{center}
    \caption{Comparison of the spatial distribution of $^{44}$Ti with that
    of other components of Cas A \citep{2017ApJ...834...19G}.  Particularly relevant is the poor correspondence with (shocked) iron.}
    \label{fig:CasA_NuSTAR_44Ti}
\end{figure}

While HEX-P can revolutionize the study of hard X-ray nonthermal continua from particle accelerators, the spectral region above 10 keV holds little information on thermal line emission.  However, HEX-P can make a major improvement in line spectroscopy of another sort: radioactive decay of freshly synthesized unstable isotopes from supernovae or from double neutron-star mergers.  In supernovae, the well-known hard X-ray nuclear decay lines at 68 and 78 keV emitted in the decay chain of $^{44}$Ti contain invaluable information on the synthesis of iron-group elements.  One of the most important results from \nustar\ was the detection of these spectral lines from SN1987A, and more significantly, their imaging as well, from Cas A.   Unlike lines from collisionally ionized iron which must be shocked to high temperatures, these lines are emitted independent of temperature, so show the presence of titanium both before and after interaction with the reverse shock in a supernova interior.  The imaging of $^{44}$Ti in Cas A violated the widely held expectation that its distribution would mirror that of $^{56}$Ni and its stable decay product $^{56}$Fe.  The famous \nustar\ image (Fig.~\ref{fig:CasA_NuSTAR_44Ti}) shows fairly poor correspondence.  However, even \nustar's 2.4 Ms integration time was not sufficient to fully characterize the distribution at lower flux levels, and the  spatial resolution, far poorer than that of the Fe K$\alpha$ image produced by Chandra, did not permit more detailed examinations of Ti/Fe ratios.    Both these 
limitations will be much less severe with HEX-P, which can improve on the \nustar\ detections, but also search for $^{44}$Ti from other young supernova remnants (both core-collapse and Type Ia, though significant titanium production in SNe Ia probably requires high degrees of explosion asymmetry).  Most excitingly, HEX-P could examine a sufficiently nearby binary neutron star merger event, with the prospect of obtaining information inaccessible to optical or soft X-ray instruments.

\subsection{$^{44}$Ti line emission from young SNRs} 

\begin{figure}[h!]
\begin{center}
\includegraphics[width=0.7 \textwidth]{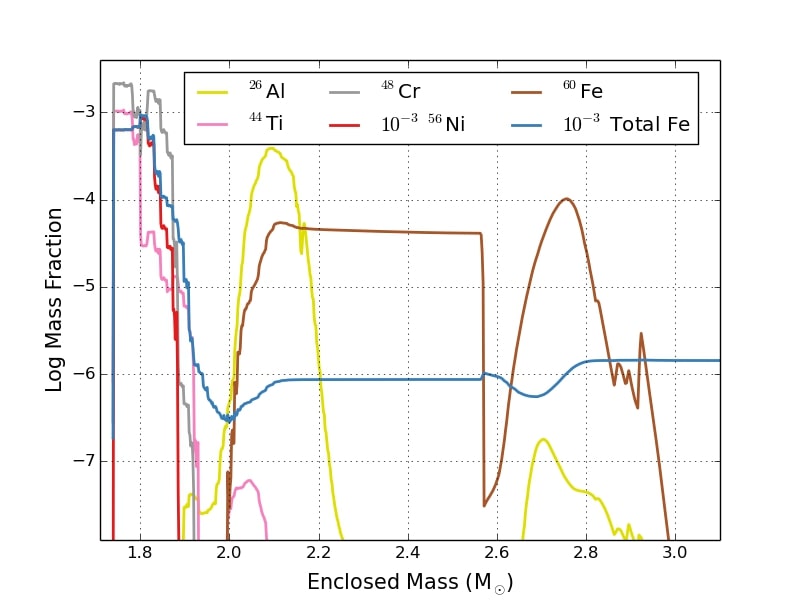}
\end{center}
\caption{Distribution of $^{26}$Al, $^{44}$Ti, $^{48}$Cr, $^{56}$Ni, $^{60}$Fe, and total Fe as a function of mass.  Like $^{56}$Ni and $^{48}$Cr, $^{44}$Ti traces the innermost ejecta.  Of the isotopes probing the innermost ejecta, $^{44}$Ti is uniquely observable in young supernova remnants.}
\label{fig:tidist}
\end{figure}

In a supernova explosion, $^{44}$Ti is produced in the innermost ejecta (either in the convective region or just above it; \cite{2010ApJS..191...66M,2023arXiv230506134F}) and is one of the most direct tracers of the supernova engine (Figure~\ref{fig:tidist}).  $^{44}$Ti production is extremely sensitive to the strength of the shock and can probe explosion asymmetries~\citep{2010ApJS..191...66M,2020ApJ...895...82V}.   But the Cas A  observations  (Fig.~\ref{fig:CasA_NuSTAR_44Ti}) highlighted issues with our understanding of the explosion mechanism.  

$^{44}$Ti provides an ideal probe of the supernova engine.  Because $^{44}$Ti is produced near the engine, it traces the explosion asymmetries caused by this engine.  \nustar\ observations of the $^{44}$Ti distribution provided the most direct evidence to date supporting the convective-engine paradigm behind core-collapse supernovae~\citep{2014Natur.506..339G,2017ApJ...834...19G,2017ApJ...842...13W,2020ApJ...895...82V}.  More precise maps of the $^{44}$Ti distribution will further constrain the asymmetries in the models.   In addition, the amount of $^{44}$Ti produced and the ratio of $^{44}$Ti to iron or other iron-peak elements provide further probes of the engine and, ultimately, the nuclear physics producing these yields.

\subsubsection{Cas A} \label{subsubsec_casa_44ti}

\begin{figure}[ht!]
  \centering
  \includegraphics[width=1.0\linewidth]{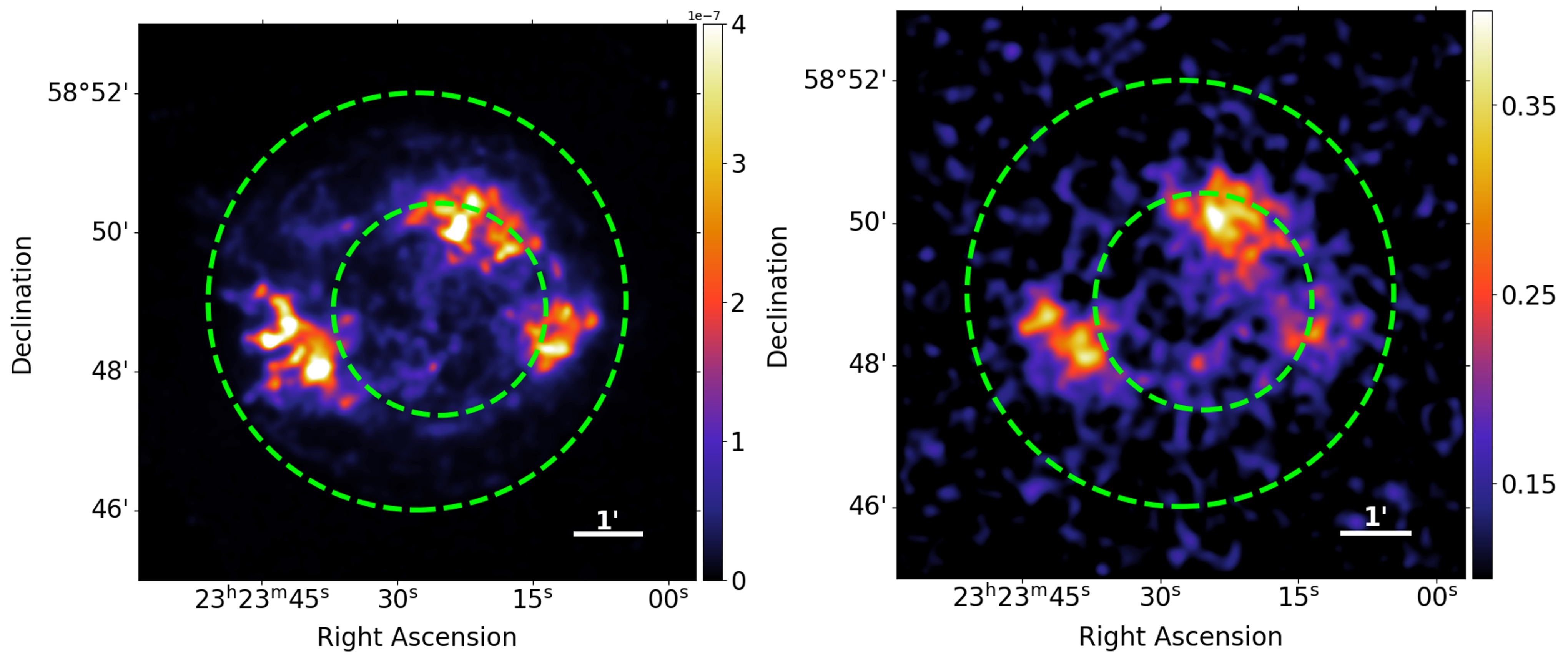}
  \caption{\textit{Left}: Chandra Fe K$\alpha$ map (6.54 -- 6.92 keV). \textit{Right}: Simulated HEX-P image of $^{44}$Ti emission from Cas A (67 -- 69 keV) for 1.5  Ms exposure using the Chandra Fe K$\alpha$ map as the input image. The green dotted lines mark the forward (outer ring) and reverse (inner ring) shock, respectively.  This image is merely to demonstrate HEX-P's high spatial
  resolution, as we have no access to the ``true" distribution of $^{44}$Ti.}
\label{fig:casa_44ti}
\end{figure}

Cas A provides the only example of spatially resolved $^{44}$Ti emission from a young SNR \citep[][Fig.~\ref{fig:CasA_NuSTAR_44Ti}]{2014Natur.506..339G,2017ApJ...834...19G}.  The distribution of the $^{44}$Ti already provides constraints on the asymmetries of the explosion.  When these observations were obtained, constraints on the iron and other iron-peak elements were limited to iron located through X-ray emission from highly ionized states, meaning that only material having passed through the reverse shock could be observed.  
The long \nustar\ observation (2.4 Ms) was limited by spatial resolution and by background; it is possible that fainter $^{44}$Ti emission could be detected and would provide quantitative measures or limits on the local Ti/Fe ratio, especially for redshifted features where both the 68 and 78 keV spectral lines could be observed.  Combining more detailed HEX-P maps with recent JWST observations of this remnant (measuring unshocked iron) will allow scientists to study details of both the shock properties and the nuclear physics. Figure \ref{fig:casa_44ti} shows the simulated HEX-P image in 67 -- 69 keV using Chandra's Fe K$\alpha$ map as a guide for the $^{44}$Ti map. We emphasize that, as Fig.~\ref{fig:CasA_NuSTAR_44Ti} graphically demonstrates \citep{2014Natur.506..339G,2017ApJ...834...19G}, the $^{44}$Ti distribution in Cas A does not trace the distribution of Fe K$\alpha$. Currently, the only available $^{44}$Ti map is from the \textit{NuSTAR} observations at a resolution of $\sim 1'$, so to exhibit the effect
of HEX-P's superior angular resolution, we use the \chandra\ Fe K$\alpha$ image as
input, knowing that the resultant simulated image is a demonstration only (compare the resolution of Fig.~\ref{fig:casa_44ti} to the \nustar\ image Fig.~\ref{fig:CasA_NuSTAR_44Ti}). However,
the integrated flux in the simulated image is that measured by \nustar.


\subsubsection{Other possible sources}


\begin{figure}[ht!]
  \centering
  \includegraphics[width=1\linewidth]{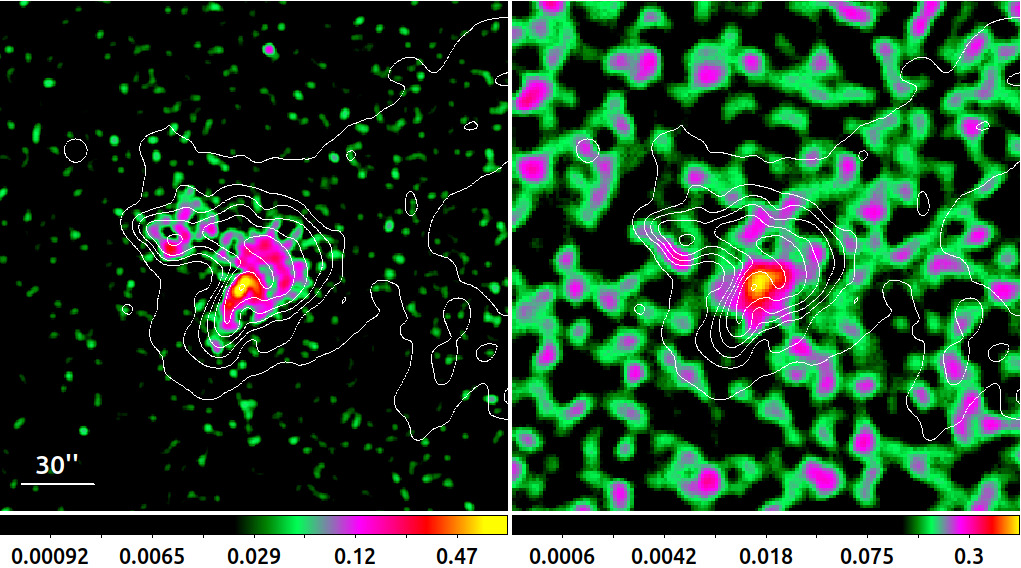}
  \caption{\textit{Left}: Chandra Fe K$\alpha$ image of G350.1$-$0.1 measured in the 6.3--6.8\,keV band. Color bar units are counts. The white contours show the Chandra continuum image. \textit{Right}: Simulated HEX-P image of $^{44}$Ti emission from G350.1$-$0.1 (67 -- 69 keV) for 2\,Ms exposure. For this simulation, the Chandra Fe K$\alpha$ map (left) was used as the input image.}
\label{fig:G350}
\end{figure}

The 85-year mean life of $^{44}$Ti means that only SNRs less than a few hundred years old will be detectable sources. Our primary target list includes (in addition to Cas A) two such objects, SN1987A, and Tycho's SNR.  SN1987A was detected by \nustar\ in a 2.6 Ms exposure, with a flux in the 68 keV line of $(3.5 \pm 0.7) \times 10^{-6}$ photons cm$^{-2}$ s$^{-1}$, at a mean epoch of 27 years after the SN.  In our proposed 300 ks HEX-P exposure, and assuming an observation year of 2029, we estimate about 10\% of the signal, 
which should allow a detection, though not an improvement over \nustar.
Type Ia supernovae such as the one that produced Tycho are not expected to synthesize large masses of $^{44}$Ti, but asymmetric explosion models have predicted values of $(1-3) \times 10^{-5}$ $M_{\odot}$ \citep{maeda10}, or about 10\% of the mass of $1.4 \times 10^{-4}$ $M_{\odot}$ inferred for Cas A \citep{2017ApJ...834...19G}.  A similar scaling of 
the flux from Cas A to the distance \citep[2.3 kpc;][]{chevalier80} and age (451 yr) of Tycho, and assuming 0.1 times the $^{44}$Ti mass of Cas A, gives a line flux of about 6\% of that of Cas A.   

In addition to Cas A, two other remnants of core-collapse supernovae have ages of order a few hundred years, as inferred from expansion measurements.  Kes 75 (G29.7-0.3), a combination shell SNR/PWN/pulsar system, has an age of $480 \pm 50$ years \citep{reynolds18}.  At a distance of
5.8 kpc, it could produce a detectable signal.  Again scaling the observed line flux from Cas A to the distance and age of Kes 75, we estimate a flux of about 7\% of that of Cas A.  Similarly, the remarkable iron-rich rapidly expanding SNR G350.1-0.3 \citep{borkowski20} is at most 600 years old (the expansion age of the fastest-moving ejecta and an upper limit to the true age, since deceleration has certainly occurred).  For an age of 600 years and a distance of 4.5\,kpc, again scaling to Cas A predicts a flux of 0.027 of that of Cas A for the 68\,keV line.  Figure~\ref{fig:G350} shows a simulation assuming (as is not the case in Cas A!) that the titanium distribution follows that of shocked iron. The strong Fe K$\alpha$ emission from G350.1-0.3 is redshifted by up to 2600 km s$^{-1}$, which if true of titanium as well, would allow both the 68 and 78 keV lines to be observed, which was not the case for Cas A, potentially doubling the relative sensitivity. The required exposure times for HEX-P detections of these targets at such flux levels are too long for the proposed primary observation program, but are certainly within reach for the extended mission.  We mention them here as a reminder that HEX-P can bring this unique capability to the study of supernovae.

\subsection{r-process nucleosynthesis in binary NS-NS mergers}

The creation of cosmic isotopes heavier than the iron group is primarily attributed to  the so-called r-process, which involves rapid neutron captures compared to $\beta$-decay
lifetimes \citep{1957RvMP...29..547B,1957PASP...69..201C}. However, it remains unclear where this nucleosynthesis occurs, as it requires neutron-rich environments with low electron fraction $Y_{\rm e}$. Recent observations of thermal kilonova emission  from ejecta produced in binary neutron-star mergers (NSMs) in gravitational wave events like GW170817 \citep{2017PhRvL.119p1101A} or in short gamma-ray bursts \citep[e.g., GRB130603B][]{2013Natur.500..547T} strongly suggest a significant r-process component.  In fact, these observations have established NSMs as a promising r-process site, possibly even more important than the standard case involving core-collapse supernovae (SNe) \citep{1974ApJ...192L.145L}.
While MeV gamma-rays from r-process nuclei should provide direct evidence for NSMs, their fluxes immediately following binary NSM events are estimated to be very low ($\sim10^{-8}$ photons s$^{-1}$ cm$^{-2}$ keV$^{-1}$) even at a close  distance of 3 Mpc \citep{2016MNRAS.459...35H}. 
This is below the sensitivity of current and near-future MeV gamma-ray missions.
An alternative and more promising approach to explore the r-process site is to search for gamma-rays from long-lived r-process nuclei in Galactic NSMs  \citep{2019ApJ...880...23W,2020ApJ...903L...3W,2020ApJ...889..168K,2022ApJ...933..111T}.
The radiation from r-process nuclei appears mainly in the sub-MeV band, but in older remnants, it can extend down to the hard X-ray band.  The hard X-ray signals from the decay of r-process isotopes can be detected by HEX-P, making it a unique probe to study the r-process from Galactic NSMs and establishing synergy with gravitational wave observations in the multi-messenger era.  

\begin{figure}[h!]
\begin{center}
\includegraphics[width=0.6 \textwidth]{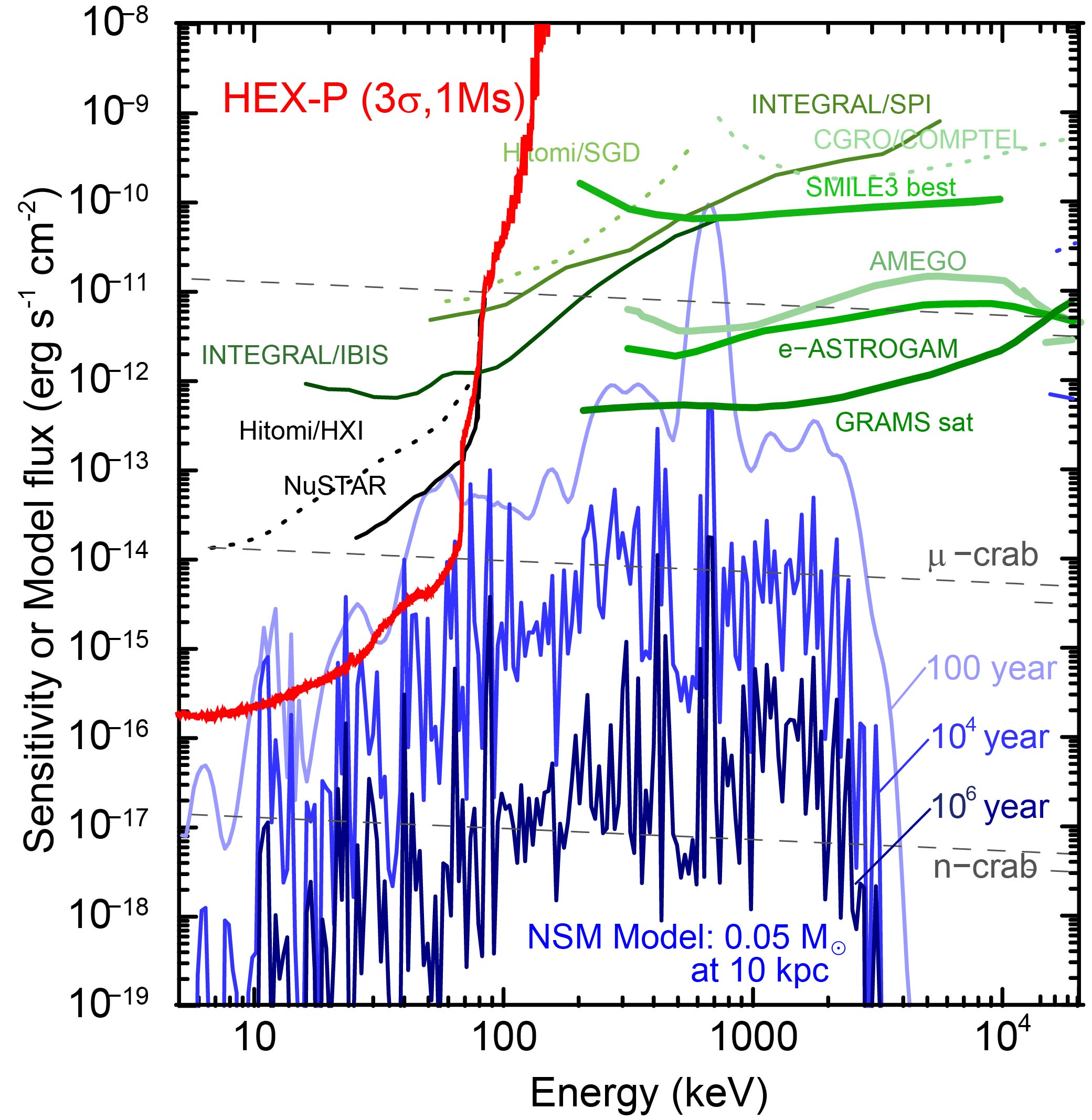}
\end{center}
\caption{Gamma-ray spectra from NSM remnants estimated by \citet{2022ApJ...933..111T} at a distance of 10 kpc and with ages of 100, $10^4$, and $10^6$ years since the merging are shown in cyan, blue, and dark blue lines, respectively. The 3-$\sigma$ sensitivities of the hard X-ray to gamma-ray missions are shown as the labeled lines. The 3-$\sigma$ sensitivity of HEX-P with 1 Ms exposure is shown in red. For others, please see Figure 14 in \citet{2022ApJ...933..111T}.}
\label{fig:NSM:sensitivity}
\end{figure}

In order to assess the feasibility of detecting gamma-rays from Galactic NSMs with HEX-P, we compared the gamma-ray spectra from Galactic NSMs, at ages of $\tau$ of 100, $10^4$, and $10^6$ years old and located at 10 kpc, with the sensitivities of hard X-ray to gamma-ray telescopes including HEX-P (Figure \ref{fig:NSM:sensitivity}). 
Since most Galactic NSMs are expected to be $\sim8$ kpc away \citep{2019ApJ...880...23W}, HEX-P can detect younger NSM remnants with $\tau$ less than a few hundred years. Additionally, HEX-P may be able to detect nuclear gamma-ray lines from  middle-aged NSM remnants ($\tau$ less than about a $\rm{few}\times10^3$ years), since the Doppler broadening caused by the fast motion of ejecta is much less significant  than in younger remnants. Figure \ref{fig:NSM:sensitivity} shows that the nuclear lines from $^{229}$Th (11.3 keV), $^{126}$Sn (23.4 keV), $^{241}$Am (26.8 keV, 59.6 keV), $^{225}$Ra (40.3 keV), and $^{243}$Am (43.9 keV) from Galactic NSMs with $10^4$ years old are above the sensitivity of HEX-P.

\begin{figure}[h!]
\begin{center}
\includegraphics[width=1.0\textwidth]{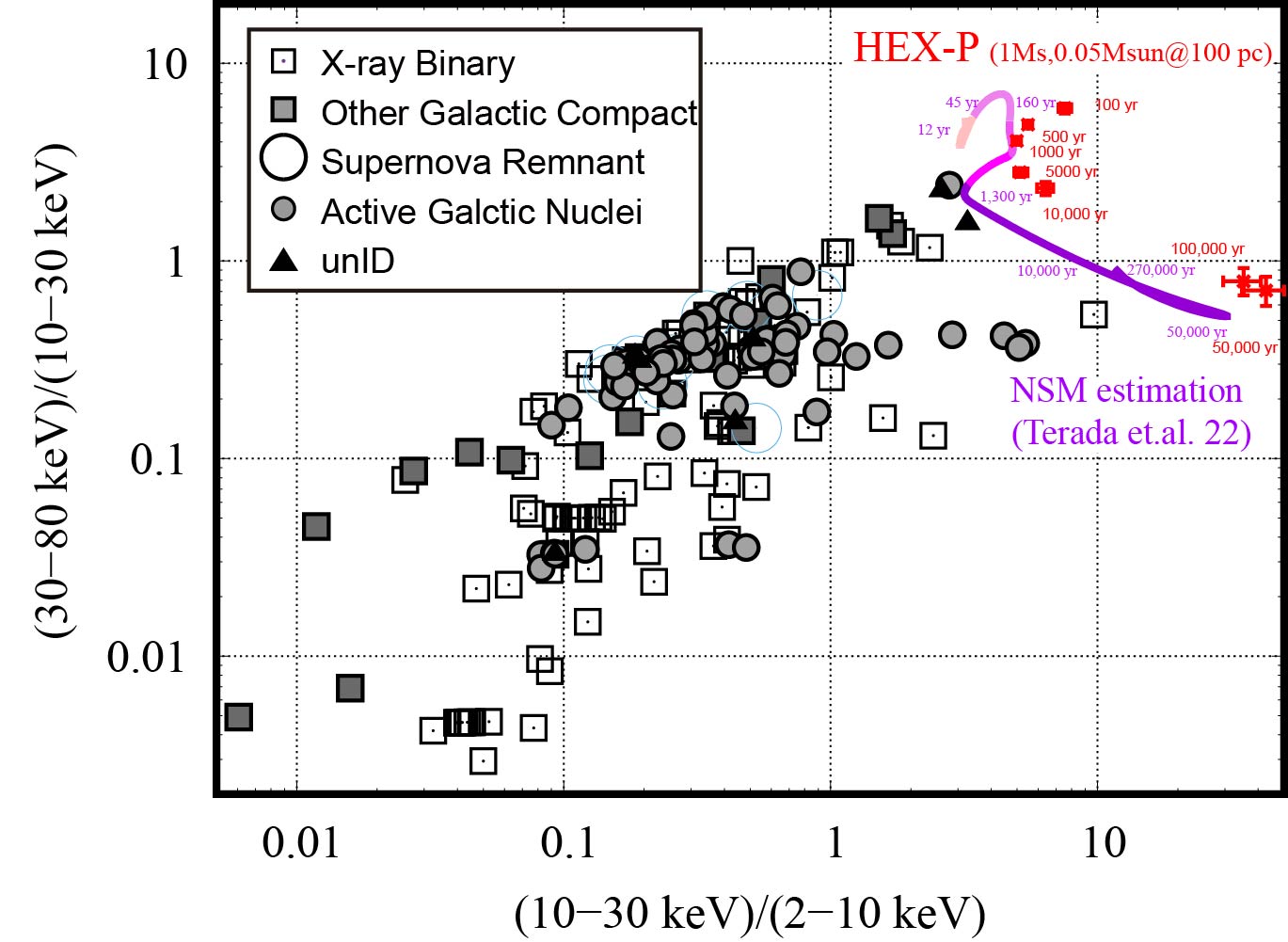}
\end{center}
\caption{A color–color diagram of the flux ratio between the 30-80 and 10-30 keV bands versus that between the 10-30 and 2-10 keV bands. Our estimation, following the same procedure as Figure 8 in \citet{2022ApJ...933..111T}, is shown in the purple line, and those of the X-ray objects listed in the \integral\ catalog (version 0043) are plotted in gray. The red crosses represent the expectation with 3$\sigma$ errors for the NSMs at 100 pc using HEX-P with 1 Ms exposure.}
\label{fig:NSM:colorcorlor}
\end{figure}

Since HEX-P is not a survey-type mission, it is important to select optimal NSM candidates in advance. This can be done through a pilot survey utilizing the current hard X-ray and gamma-ray missions, such as \integral, \swift, {\it MAXI}, etc, or by discovering NSMs serendipitously, for instance in the HEX-P's Galactic Center survey (Mori et al., submitted to FrASS). One of the effective ways based on a pilot survey with the current gamma-ray instruments is to diagnose the spectral shape of NSMs using the color-color diagram shown in Figure 8 in \citet{2022ApJ...933..111T}. 
Figure \ref{fig:NSM:colorcorlor} displays the same diagram, but recalculated for HEX-P's energy band from 2 to 80 keV, using the simulation code adopted from \citet{ 2022ApJ...933..111T}. Our simulations demonstrate that very young NSMs ($\tau< 10^3$ y) and old NSMs ($\tau > 10^4$ y) can be distinguished from other types of hard X-ray sources such as SNRs, AGNs, and XRBs.  
The figure also demonstrates the numerical results for the case of a nearby NSM (within 100 pc) observed by HEX-P with long exposure times (shown in red crosses in Figure \ref{fig:NSM:colorcorlor}).

Once an NSM remnant is identified by the color-color diagram analysis, a handful of nuclear lines from r-process nuclei may be detected with deeper observations of HEX-P, as already shown in Figure \ref{fig:NSM:sensitivity}. Nuclear lines in the hard X-ray band originate mostly from nuclei in the relatively low $Y_{\rm e}$ environment, compared with those in the MeV band. Note that the hard X-ray nuclear lines are unique signatures of NSMs, while MeV gamma-ray lines could be emitted from SNe. 
Therefore, the hard X-ray lines potentially detectable with HEX-P, namely $^{229}$Th, $^{126}$Sn, $^{241}$Am, $^{225}$Ra, and $^{243}$Am, will provide strong evidence of r-process nucleosynthesis in NSM remnants. 
Among them, it is anticipated that the nuclear lines from $^{241}$Am, $^{243}$Am, and $^{126}$Sn 
 have constant luminosities for  $\tau \sim 10^2$ to $10^5$ years (see Figure 9 in \cite{2022ApJ...933..111T}), yielding  a few hundred photon counts with 1 Msec exposure, and therefore, making them suitable as a standard candle for estimating the source distance. 
Overall, searching for r-process nucleosynthesis sites in our Galaxy is a unique scientific objective of HEX-P in the 2030s that, while somewhat risky, has the potential for significant scientific gains.



\section{Conclusions} 

The physical processes involved in the acceleration of particles at
strong shocks are particularly well exhibited in shell SNRs, for
nonrelativistic shocks, and in PWNe, for relativistic shocks.  These
processes ought to operate across the Universe, in many environments
where sources are too compact to be spatially resolved.  Galactic
SNRs and PWNe, closely examined by HEX-P, can serve as laboratories
in which to investigate details of these processes.   Major breakthroughs in understanding occurred with hard X-ray observations
of the famous shell SNR SN 1006, and of the even more famous
Crab Nebula.  A great deal more can be learned about objects
in each class with the new capabilities promised by HEX-P.  The simulations described above demonstrate the kinds of advances possible in the study of particular SNRs and PWNe, and of Galactic particle acceleration in general (see Mori et al. 2023 for other types of particle accelerators).  The additional, unique capability of HEX-P to study nuclear-decay line
emission will improve our knowledge of the $^{44}$Ti distribution in Cas A, and may
detect lines from other young SNRs or in the unlikely but exciting possibility of a 
sufficiently nearby neutron-star binary merger event.
HEX-P clearly stands out as the foremost X-ray observatory for cosmic accelerators and nuclear astrophysics in the 2030s.

\section*{Author Contributions}

\S 1 (Reynolds).  \S 2 (Madsen and Garcia). \S 3 (Reynolds). \S 3.3 (Woo). \S 3.4 (Abdelmaguid, Alford). \S3.5 (Reynolds, Abdelmaguid). \S 3.6 (Mori). \S 4 (An, Park, Kim). \S 4.4 (Nynka, Abdelmaguid). \S 5 (Reynolds). \S 5.1 (Fryer, Reynolds, Woo, Alford, Park). \S 5.2 (Terada, Motogami, Ohsumi). \S 6 (Reynolds).   



\section*{Funding}
The work of D.S. was carried out at the Jet Propulsion Laboratory, California Institute of Technology, under a contract with NASA.

\section*{Acknowledgments} 
We are grateful to J. Wilms, T. Dauser, C. Kirsch, M. Lorenz, L. Dauner, and the SIXTE development team for their assistance with SIXTE simulations.  We thank K. Fang, L. Lu, and K. Malone for assistance.
HA acknowledges support from the National Research Foundation of Korea (NRF) grant funded by the Korean Government (MSIT) (NRF-2023R1A2C1002718).
This work was financially supported by Japan Society for the Promotion of Science Grants-in-Aid for Scientific Research (KAKENHI) Grant Numbers, JP23H01211 (AB),
22K14064 (NT), 20K04009 (YT).



\bibliographystyle{Frontiers-Harvard} 


\bibliography{main}

\begin{thebibliography}{136}
\providecommand{\natexlab}[1]{#1}
\expandafter\ifx\csname urlstyle\endcsname\relax
  \providecommand{\doi}[1]{doi:\discretionary{}{}{}#1}\else
  \providecommand{\doi}{doi:\discretionary{}{}{}\begingroup
  \urlstyle{rm}\Url}\fi
\providecommand{\selectlanguage}[1]{\relax}
\providecommand{\bibAnnoteFile}[1]{%
  \IfFileExists{#1}{\begin{quotation}\noindent\textsc{Key:} #1\\
  \textsc{Annotation:}\ \input{#1}\end{quotation}}{}}
\providecommand{\bibAnnote}[2]{%
  \begin{quotation}\noindent\textsc{Key:} #1\\
  \textsc{Annotation:}\ #2\end{quotation}}

\bibitem[{{Abbott} et~al.(2017){Abbott}, {Abbott}, {Abbott}, {Acernese},
  {Ackley}, {Adams} et~al.}]{2017PhRvL.119p1101A}
{Abbott}, B.~P., {Abbott}, R., {Abbott}, T.~D., {Acernese}, F., {Ackley}, K.,
  {Adams}, C., et~al. (2017).
\newblock {GW170817: Observation of Gravitational Waves from a Binary Neutron
  Star Inspiral}.
\newblock \emph{\prl} 119, 161101.
\newblock \doi{10.1103/PhysRevLett.119.161101}
\bibAnnoteFile{2017PhRvL.119p1101A}

\bibitem[{{Abdelmaguid} et~al.(2023){Abdelmaguid}, {Gelfand}, {Gotthelf}, and
  {Straal}}]{Abdelmaguid2023}
{Abdelmaguid}, M., {Gelfand}, J.~D., {Gotthelf}, E., and {Straal}, S. (2023).
\newblock {Broadband X-Ray Spectroscopy of the Pulsar Wind Nebula in HESS
  J1640-465}.
\newblock \emph{\apj} 946, 40.
\newblock \doi{10.3847/1538-4357/acbd30}
\bibAnnoteFile{Abdelmaguid2023}

\bibitem[{{Abeysekara} et~al.(2017{\natexlab{a}}){Abeysekara}, {Albert},
  {Alfaro}, {Alvarez}, {{\'A}lvarez}, {Arceo} et~al.}]{abeysekara2017}
{Abeysekara}, A.~U., {Albert}, A., {Alfaro}, R., {Alvarez}, C., {{\'A}lvarez},
  J.~D., {Arceo}, R., et~al. (2017{\natexlab{a}}).
\newblock {Extended gamma-ray sources around pulsars constrain the origin of
  the positron flux at Earth}.
\newblock \emph{Science} 358, 911--914.
\newblock \doi{10.1126/science.aan4880}
\bibAnnoteFile{abeysekara2017}

\bibitem[{{Abeysekara} et~al.(2017{\natexlab{b}}){Abeysekara}, {Albert},
  {Alfaro}, {Alvarez}, {{\'A}lvarez}, {Arceo} et~al.}]{Abeysekara2017b}
{Abeysekara}, A.~U., {Albert}, A., {Alfaro}, R., {Alvarez}, C., {{\'A}lvarez},
  J.~D., {Arceo}, R., et~al. (2017{\natexlab{b}}).
\newblock {Extended gamma-ray sources around pulsars constrain the origin of
  the positron flux at Earth}.
\newblock \emph{Science} 358, 911--914.
\newblock \doi{10.1126/science.aan4880}
\bibAnnoteFile{Abeysekara2017b}

\bibitem[{{Acciari} et~al.(2011){Acciari}, {Aliu}, {Arlen}, {Aune}, {Beilicke},
  {Benbow} et~al.}]{2011ApJ...730L..20A}
{Acciari}, V.~A., {Aliu}, E., {Arlen}, T., {Aune}, T., {Beilicke}, M.,
  {Benbow}, W., et~al. (2011).
\newblock {Discovery of TeV Gamma-ray Emission from Tycho's Supernova Remnant}.
\newblock \emph{\apjl} 730, L20.
\newblock \doi{10.1088/2041-8205/730/2/L20}
\bibAnnoteFile{2011ApJ...730L..20A}

\bibitem[{{Adams} et~al.(2021){Adams}, {Benbow}, {Brill}, {Brose},
  {Buchovecky}, {Capasso} et~al.}]{Adams2021a}
{Adams}, C.~B., {Benbow}, W., {Brill}, A., {Brose}, R., {Buchovecky}, M.,
  {Capasso}, M., et~al. (2021).
\newblock {VERITAS Observations of the Galactic Center Region at Multi-TeV
  Gamma-Ray Energies}.
\newblock \emph{\apj} 913, 115.
\newblock \doi{10.3847/1538-4357/abf926}
\bibAnnoteFile{Adams2021a}

\bibitem[{{Aguilar} et~al.(2019){Aguilar}, {Ali Cavasonza}, {Alpat}, {Ambrosi},
  {Arruda}, {Attig} et~al.}]{Aguilar2019}
{Aguilar}, M., {Ali Cavasonza}, L., {Alpat}, B., {Ambrosi}, G., {Arruda}, L.,
  {Attig}, N., et~al. (2019).
\newblock {Towards Understanding the Origin of Cosmic-Ray Electrons}.
\newblock \emph{\prl} 122, 101101.
\newblock \doi{10.1103/PhysRevLett.122.101101}
\bibAnnoteFile{Aguilar2019}

\bibitem[{{Aharonian} et~al.(2005){Aharonian}, {Akhperjanian}, {Aye},
  {Bazer-Bachi}, {Beilicke}, {Benbow} et~al.}]{Aharonian2005}
{Aharonian}, F., {Akhperjanian}, A.~G., {Aye}, K.-M., {Bazer-Bachi}, A.~R.,
  {Beilicke}, M., {Benbow}, W., et~al. (2005).
\newblock {Very high energy gamma rays from the composite SNR G 0.9+0.1}.
\newblock \emph{\aap} 432, L25--L29.
\newblock \doi{10.1051/0004-6361:200500022}
\bibAnnoteFile{Aharonian2005}

\bibitem[{{Alp} et~al.(2021){Alp}, {Larsson}, and {Fransson}}]{Alp2021}
{Alp}, D., {Larsson}, J., and {Fransson}, C. (2021).
\newblock {Thermal Emission and Radioactive Lines, but No Pulsar, in the
  Broadband X-Ray Spectrum of Supernova 1987A}.
\newblock \emph{\apj} 916, 76.
\newblock \doi{10.3847/1538-4357/ac052d}
\bibAnnoteFile{Alp2021}

\bibitem[{{Alp} et~al.(2018){Alp}, {Larsson}, {Fransson}, {Indebetouw},
  {Jerkstrand}, {Ahola} et~al.}]{Alp2018}
{Alp}, D., {Larsson}, J., {Fransson}, C., {Indebetouw}, R., {Jerkstrand}, A.,
  {Ahola}, A., et~al. (2018).
\newblock {The 30 Year Search for the Compact Object in SN 1987A}.
\newblock \emph{\apj} 864, 174.
\newblock \doi{10.3847/1538-4357/aad739}
\bibAnnoteFile{Alp2018}

\bibitem[{{An}(2019)}]{An2019}
{An}, H. (2019).
\newblock {NuSTAR Hard X-Ray Studies of the Pulsar Wind Nebula 3C 58}.
\newblock \emph{\apj} 876, 150.
\newblock \doi{10.3847/1538-4357/ab18a6}
\bibAnnoteFile{An2019}

\bibitem[{{An} et~al.(2014){An}, {Madsen}, {Reynolds}, {Kaspi}, {Harrison},
  {Boggs} et~al.}]{An2014}
{An}, H., {Madsen}, K.~K., {Reynolds}, S.~P., {Kaspi}, V.~M., {Harrison},
  F.~A., {Boggs}, S.~E., et~al. (2014).
\newblock {High-energy X-Ray Imaging of the Pulsar Wind Nebula MSH 15-52:
  Constraints on Particle Acceleration and Transport}.
\newblock \emph{\apj} 793, 90.
\newblock \doi{10.1088/0004-637X/793/2/90}
\bibAnnoteFile{An2014}

\bibitem[{{Arnaud}(1996)}]{Arnaud1996}
{Arnaud}, K.~A. (1996).
\newblock {XSPEC: The First Ten Years}.
\newblock In \emph{Astronomical Data Analysis Software and Systems V}, eds.
  G.~H. {Jacoby} and J.~{Barnes}. vol. 101 of \emph{Astronomical Society of the
  Pacific Conference Series}, 17
\bibAnnoteFile{Arnaud1996}

\bibitem[{{Atoyan} and {Dermer}(2012)}]{2012ApJ...749L..26A}
{Atoyan}, A. and {Dermer}, C.~D. (2012).
\newblock {Gamma Rays from the Tycho Supernova Remnant: Multi-zone versus
  Single-zone Modeling}.
\newblock \emph{\apjl} 749, L26.
\newblock \doi{10.1088/2041-8205/749/2/L26}
\bibAnnoteFile{2012ApJ...749L..26A}

\bibitem[{{Bamba} et~al.(2010){Bamba}, {Anada}, {Dotani}, {Mori}, {Yamazaki},
  {Ebisawa} et~al.}]{bamba10}
{Bamba}, A., {Anada}, T., {Dotani}, T., {Mori}, K., {Yamazaki}, R., {Ebisawa},
  K., et~al. (2010).
\newblock {X-ray Evolution of Pulsar Wind Nebulae}.
\newblock \emph{\apjl} 719, L116--L120.
\newblock \doi{10.1088/2041-8205/719/2/L116}
\bibAnnoteFile{bamba10}

\bibitem[{{Bamba} et~al.(2022){Bamba}, {Shibata}, {Tanaka}, {Mori}, {Uchida},
  {Terada} et~al.}]{bamba22}
{Bamba}, A., {Shibata}, S., {Tanaka}, S.~J., {Mori}, K., {Uchida}, H.,
  {Terada}, Y., et~al. (2022).
\newblock {Spectral break of energetic pulsar wind nebulae detected with
  wideband X-ray observations}.
\newblock \emph{\pasj} 74, 1186--1197.
\newblock \doi{10.1093/pasj/psac062}
\bibAnnoteFile{bamba22}

\bibitem[{{Bamba} et~al.(2005){Bamba}, {Yamazaki}, {Yoshida}, {Terasawa}, and
  {Koyama}}]{2005ApJ...621..793B}
{Bamba}, A., {Yamazaki}, R., {Yoshida}, T., {Terasawa}, T., and {Koyama}, K.
  (2005).
\newblock {A Spatial and Spectral Study of Nonthermal Filaments in Historical
  Supernova Remnants: Observational Results with Chandra}.
\newblock \emph{\apj} 621, 793--802.
\newblock \doi{10.1086/427620}
\bibAnnoteFile{2005ApJ...621..793B}

\bibitem[{{Bandiera}(2008)}]{Bandiera2008}
{Bandiera}, R. (2008).
\newblock {On the X-ray feature associated with the Guitar nebula}.
\newblock \emph{\aap} 490, L3--L6.
\newblock \doi{10.1051/0004-6361:200810666}
\bibAnnoteFile{Bandiera2008}

\bibitem[{{Bandiera} et~al.(2020){Bandiera}, {Bucciantini}, {Mart{\'\i}n},
  {Olmi}, and {Torres}}]{Bandiera2020}
{Bandiera}, R., {Bucciantini}, N., {Mart{\'\i}n}, J., {Olmi}, B., and {Torres},
  D.~F. (2020).
\newblock {Reverberation of pulsar wind nebulae (I): impact of the medium
  properties and other parameters upon the extent of the compression}.
\newblock \emph{\mnras} 499, 2051--2062.
\newblock \doi{10.1093/mnras/staa2956}
\bibAnnoteFile{Bandiera2020}

\bibitem[{{Bandiera} et~al.(2023){Bandiera}, {Bucciantini}, {Mart{\'\i}n},
  {Olmi}, and {Torres}}]{Bandiera2023}
{Bandiera}, R., {Bucciantini}, N., {Mart{\'\i}n}, J., {Olmi}, B., and {Torres},
  D.~F. (2023).
\newblock {Reverberation of pulsar wind nebulae - II. Anatomy of the
  'thin-shell' evolution}.
\newblock \emph{\mnras} 520, 2451--2472.
\newblock \doi{10.1093/mnras/stad134}
\bibAnnoteFile{Bandiera2023}

\bibitem[{{Berezhko} et~al.(2013){Berezhko}, {Ksenofontov}, and
  {V{\"o}lk}}]{2013ApJ...763...14B}
{Berezhko}, E.~G., {Ksenofontov}, L.~T., and {V{\"o}lk}, H.~J. (2013).
\newblock {The Nature of Gamma-Ray Emission of Tycho's Supernova Remnant}.
\newblock \emph{\apj} 763, 14.
\newblock \doi{10.1088/0004-637X/763/1/14}
\bibAnnoteFile{2013ApJ...763...14B}

\bibitem[{{Bleeker} et~al.(2001){Bleeker}, {Willingale}, {van der Heyden},
  {Dennerl}, {Kaastra}, {Aschenbach} et~al.}]{2001A&A...365L.225B}
{Bleeker}, J.~A.~M., {Willingale}, R., {van der Heyden}, K., {Dennerl}, K.,
  {Kaastra}, J.~S., {Aschenbach}, B., et~al. (2001).
\newblock {Cassiopeia A: On the origin of the hard X-ray continuum and the
  implication of the observed O Viii Ly-alpha /Ly-beta distribution}.
\newblock \emph{\aap} 365, L225--L230.
\newblock \doi{10.1051/0004-6361:20000048}
\bibAnnoteFile{2001A&A...365L.225B}

\bibitem[{{Borkowski} et~al.(2020){Borkowski}, {Miltich}, and
  {Reynolds}}]{borkowski20}
{Borkowski}, K.~J., {Miltich}, W., and {Reynolds}, S.~P. (2020).
\newblock {Expansion and Age of the Supernova Remnant G350.1-0.3: High-velocity
  Iron Ejecta from a Core-collapse Event}.
\newblock \emph{\apjl} 905, L19.
\newblock \doi{10.3847/2041-8213/abcda7}
\bibAnnoteFile{borkowski20}

\bibitem[{{Borkowski} et~al.(2010){Borkowski}, {Reynolds}, {Green}, {Hwang},
  {Petre}, {Krishnamurthy} et~al.}]{borkowski10}
{Borkowski}, K.~J., {Reynolds}, S.~P., {Green}, D.~A., {Hwang}, U., {Petre},
  R., {Krishnamurthy}, K., et~al. (2010).
\newblock {Radioactive Scandium in the Youngest Galactic Supernova Remnant
  G1.9+0.3}.
\newblock \emph{\apjl} 724, L161--L165.
\newblock \doi{10.1088/2041-8205/724/2/L161}
\bibAnnoteFile{borkowski10}

\bibitem[{{Bucciantini} et~al.(2003){Bucciantini}, {Blondin}, {Del Zanna}, and
  {Amato}}]{Bucciantini2003}
{Bucciantini}, N., {Blondin}, J.~M., {Del Zanna}, L., and {Amato}, E. (2003).
\newblock {Spherically symmetric relativistic MHD simulations of pulsar wind
  nebulae in supernova remnants}.
\newblock \emph{\aap} 405, 617--626.
\newblock \doi{10.1051/0004-6361:20030624}
\bibAnnoteFile{Bucciantini2003}

\bibitem[{{Burbidge} et~al.(1957){Burbidge}, {Burbidge}, {Fowler}, and
  {Hoyle}}]{1957RvMP...29..547B}
{Burbidge}, E.~M., {Burbidge}, G.~R., {Fowler}, W.~A., and {Hoyle}, F. (1957).
\newblock {Synthesis of the Elements in Stars}.
\newblock \emph{Reviews of Modern Physics} 29, 547--650.
\newblock \doi{10.1103/RevModPhys.29.547}
\bibAnnoteFile{1957RvMP...29..547B}

\bibitem[{{Burgess} et~al.(2022){Burgess}, {Mori}, {Gelfand}, {Hailey},
  {Tokayer}, {Woo} et~al.}]{Burgess2022}
{Burgess}, D.~A., {Mori}, K., {Gelfand}, J.~D., {Hailey}, C.~J., {Tokayer},
  Y.~M., {Woo}, J., et~al. (2022).
\newblock {The Eel Pulsar Wind Nebula: A PeVatron-candidate Origin for HAWC
  J1826-128 and HESS J1826-130}.
\newblock \emph{\apj} 930, 148.
\newblock \doi{10.3847/1538-4357/ac650a}
\bibAnnoteFile{Burgess2022}

\bibitem[{{Cameron}(1957)}]{1957PASP...69..201C}
{Cameron}, A.~G.~W. (1957).
\newblock {Nuclear Reactions in Stars and Nucleogenesis}.
\newblock \emph{\pasp} 69, 201.
\newblock \doi{10.1086/127051}
\bibAnnoteFile{1957PASP...69..201C}

\bibitem[{{Camilo} et~al.(2009){Camilo}, {Ransom}, {Gaensler}, and
  {Lorimer}}]{camilo09}
{Camilo}, F., {Ransom}, S.~M., {Gaensler}, B.~M., and {Lorimer}, D.~R. (2009).
\newblock {Discovery of the Energetic Pulsar J1747-2809 in the Supernova
  Remnant G0.9+0.1}.
\newblock \emph{\apjl} 700, L34--L38.
\newblock \doi{10.1088/0004-637X/700/1/L34}
\bibAnnoteFile{camilo09}

\bibitem[{{Cao} et~al.(2023){Cao}, {Aharonian}, {An}, {Axikegu}, {Bai}, {Bao}
  et~al.}]{lhasso23}
{Cao}, Z., {Aharonian}, F., {An}, Q., {Axikegu}, {Bai}, Y.~X., {Bao}, Y.~W.,
  et~al. (2023).
\newblock {The First LHAASO Catalog of Gamma-Ray Sources}.
\newblock \emph{arXiv e-prints} ,
  arXiv:2305.17030\doi{10.48550/arXiv.2305.17030}
\bibAnnoteFile{lhasso23}

\bibitem[{{Cao} et~al.(2021){Cao}, {Aharonian}, {An}, {Axikegu}, {Bai}, {Bao}
  et~al.}]{Cao2021}
{Cao}, Z., {Aharonian}, F.~A., {An}, Q., {Axikegu}, L.~X., Bai, {Bai}, Y.~X.,
  {Bao}, Y.~W., et~al. (2021).
\newblock {Ultrahigh-energy photons up to 1.4 petaelectronvolts from 12
  {\ensuremath{\gamma}}-ray Galactic sources}.
\newblock \emph{\nat} 594, 33--36.
\newblock \doi{10.1038/s41586-021-03498-z}
\bibAnnoteFile{Cao2021}

\bibitem[{{Carlton} et~al.(2011){Carlton}, {Borkowski}, {Reynolds}, {Hwang},
  {Petre}, {Green} et~al.}]{carlton11}
{Carlton}, A.~K., {Borkowski}, K.~J., {Reynolds}, S.~P., {Hwang}, U., {Petre},
  R., {Green}, D.~A., et~al. (2011).
\newblock {Expansion of the Youngest Galactic Supernova Remnant G1.9+0.3}.
\newblock \emph{\apjl} 737, L22.
\newblock \doi{10.1088/2041-8205/737/1/L22}
\bibAnnoteFile{carlton11}

\bibitem[{{Cherenkov Telescope Array Consortium}(2023)}]{CTA2023}
{Cherenkov Telescope Array Consortium}, T. (2023).
\newblock {Sensitivity of the Cherenkov Telescope Array to TeV photon emission
  from the Large Magellanic Cloud}.
\newblock \emph{arXiv e-prints} ,
  arXiv:2305.16707\doi{10.48550/arXiv.2305.16707}
\bibAnnoteFile{CTA2023}

\bibitem[{{Chevalier} et~al.(1980){Chevalier}, {Kirshner}, and
  {Raymond}}]{chevalier80}
{Chevalier}, R.~A., {Kirshner}, R.~P., and {Raymond}, J.~C. (1980).
\newblock {The optical emission from a fast shock wave with application to
  supernova remnants.}
\newblock \emph{\apj} 235, 186--195.
\newblock \doi{10.1086/157623}
\bibAnnoteFile{chevalier80}

\bibitem[{{Cholis} and {Hooper}(2013)}]{Cholis2013}
{Cholis}, I. and {Hooper}, D. (2013).
\newblock {Dark matter and pulsar origins of the rising cosmic ray positron
  fraction in light of new data from the AMS}.
\newblock \emph{\prd} 88, 023013.
\newblock \doi{10.1103/PhysRevD.88.023013}
\bibAnnoteFile{Cholis2013}

\bibitem[{{Cigan} et~al.(2019){Cigan}, {Matsuura}, {Gomez}, {Indebetouw},
  {Abell{\'a}n}, {Gabler} et~al.}]{Cigan2019}
{Cigan}, P., {Matsuura}, M., {Gomez}, H.~L., {Indebetouw}, R., {Abell{\'a}n},
  F., {Gabler}, M., et~al. (2019).
\newblock {High Angular Resolution ALMA Images of Dust and Molecules in the SN
  1987A Ejecta}.
\newblock \emph{\apj} 886, 51.
\newblock \doi{10.3847/1538-4357/ab4b46}
\bibAnnoteFile{Cigan2019}

\bibitem[{{Dauser} et~al.(2019){Dauser}, {Falkner}, {Lorenz}, {Kirsch},
  {Peille}, {Cucchetti} et~al.}]{Dauser2019}
{Dauser}, T., {Falkner}, S., {Lorenz}, M., {Kirsch}, C., {Peille}, P.,
  {Cucchetti}, E., et~al. (2019).
\newblock {SIXTE: a generic X-ray instrument simulation toolkit}.
\newblock \emph{\aap} 630, A66.
\newblock \doi{10.1051/0004-6361/201935978}
\bibAnnoteFile{Dauser2019}

\bibitem[{{de O{\~n}a Wilhelmi} et~al.(2022){de O{\~n}a Wilhelmi},
  {L{\'o}pez-Coto}, {Amato}, and {Aharonian}}]{Wilhelmi2022}
{de O{\~n}a Wilhelmi}, E., {L{\'o}pez-Coto}, R., {Amato}, E., and {Aharonian},
  F. (2022).
\newblock {On the Potential of Bright, Young Pulsars to Power Ultrahigh
  Gamma-Ray Sources}.
\newblock \emph{\apjl} 930, L2.
\newblock \doi{10.3847/2041-8213/ac66cf}
\bibAnnoteFile{Wilhelmi2022}

\bibitem[{{Del Zanna} et~al.(2004){Del Zanna}, {Amato}, and
  {Bucciantini}}]{delzanna2004}
{Del Zanna}, L., {Amato}, E., and {Bucciantini}, N. (2004).
\newblock {Axially symmetric relativistic MHD simulations of Pulsar Wind
  Nebulae in Supernova Remnants. On the origin of torus and jet-like features}.
\newblock \emph{\aap} 421, 1063--1073.
\newblock \doi{10.1051/0004-6361:20035936}
\bibAnnoteFile{delzanna2004}

\bibitem[{{Della Torre} et~al.(2015){Della Torre}, {Gervasi}, {Rancoita},
  {Rozza}, and {Treves}}]{DellaTorre2015}
{Della Torre}, S., {Gervasi}, M., {Rancoita}, P.~G., {Rozza}, D., and {Treves},
  A. (2015).
\newblock {Pulsar Wind Nebulae as a source of the observed electron and
  positron excess at high energy: The case of Vela-X}.
\newblock \emph{Journal of High Energy Astrophysics} 8, 27--34.
\newblock \doi{10.1016/j.jheap.2015.08.001}
\bibAnnoteFile{DellaTorre2015}

\bibitem[{{Diesing} and {Caprioli}(2021)}]{Diesing2021}
{Diesing}, R. and {Caprioli}, D. (2021).
\newblock {Steep Cosmic-Ray Spectra with Revised Diffusive Shock Acceleration}.
\newblock \emph{\apj} 922, 1.
\newblock \doi{10.3847/1538-4357/ac22fe}
\bibAnnoteFile{Diesing2021}

\bibitem[{{Eraerds} et~al.(2021){Eraerds}, {Antonelli}, {Davis}, {Hall},
  {Hetherington}, {Holland} et~al.}]{Eraerds2021}
{Eraerds}, T., {Antonelli}, V., {Davis}, C., {Hall}, D., {Hetherington}, O.,
  {Holland}, A., et~al. (2021).
\newblock {Enhanced simulations on the Athena/Wide Field Imager instrumental
  background}.
\newblock \emph{Journal of Astronomical Telescopes, Instruments, and Systems}
  7, 034001.
\newblock \doi{10.1117/1.JATIS.7.3.034001}
\bibAnnoteFile{Eraerds2021}

\bibitem[{{Eriksen} et~al.(2011){Eriksen}, {Hughes}, {Badenes}, {Fesen},
  {Ghavamian}, {Moffett} et~al.}]{2011ApJ...728L..28E}
{Eriksen}, K.~A., {Hughes}, J.~P., {Badenes}, C., {Fesen}, R., {Ghavamian}, P.,
  {Moffett}, D., et~al. (2011).
\newblock {Evidence for Particle Acceleration to the Knee of the Cosmic Ray
  Spectrum in Tycho's Supernova Remnant}.
\newblock \emph{\apjl} 728, L28.
\newblock \doi{10.1088/2041-8205/728/2/L28}
\bibAnnoteFile{2011ApJ...728L..28E}

\bibitem[{{Fiori} et~al.(2020){Fiori}, {Zampieri}, {Burtovoi}, {Caraveo}, and
  {Tibaldo}}]{Fiori2020}
{Fiori}, M., {Zampieri}, L., {Burtovoi}, A., {Caraveo}, P., and {Tibaldo}, L.
  (2020).
\newblock {Constraining models of the pulsar wind nebula in SNR G0.9+0.1 via
  simulation of its detection properties using the Cherenkov Telescope Array}.
\newblock \emph{\mnras} 499, 3494--3509.
\newblock \doi{10.1093/mnras/staa3039}
\bibAnnoteFile{Fiori2020}

\bibitem[{{Frank} et~al.(2016){Frank}, {Zhekov}, {Park}, {McCray}, {Dwek}, and
  {Burrows}}]{Frank2016}
{Frank}, K.~A., {Zhekov}, S.~A., {Park}, S., {McCray}, R., {Dwek}, E., and
  {Burrows}, D.~N. (2016).
\newblock {Chandra Observes the End of an Era in SN 1987A}.
\newblock \emph{\apj} 829, 40.
\newblock \doi{10.3847/0004-637X/829/1/40}
\bibAnnoteFile{Frank2016}

\bibitem[{Fraschetti et~al.(2018)Fraschetti, Katsuda, Sato, Jokipii, and
  Giacalone}]{CasA_inward}
Fraschetti, F., Katsuda, S., Sato, T., Jokipii, J.~R., and Giacalone, J.
  (2018).
\newblock Vortical amplification of the magnetic field at an inward shock of
  supernova remnant cassiopeia a.
\newblock \emph{Phys. Rev. Lett.} 120, 251101.
\newblock \doi{10.1103/PhysRevLett.120.251101}
\bibAnnoteFile{CasA_inward}

\bibitem[{{Fryer} et~al.(2023){Fryer}, {Burns}, {Hungerford}, {Safi-Harb},
  {Wollaeger}, {Miller} et~al.}]{2023arXiv230506134F}
{Fryer}, C.~L., {Burns}, E., {Hungerford}, A., {Safi-Harb}, S., {Wollaeger},
  R.~T., {Miller}, R.~S., et~al. (2023).
\newblock {Multi-Messenger Diagnostics of the Engine behind Core-Collapse
  Supernovae}.
\newblock \emph{arXiv e-prints} ,
  arXiv:2305.06134\doi{10.48550/arXiv.2305.06134}
\bibAnnoteFile{2023arXiv230506134F}

\bibitem[{{Gaensler} et~al.(2001){Gaensler}, {Pivovaroff}, and
  {Garmire}}]{Gaensler2001}
{Gaensler}, B.~M., {Pivovaroff}, M.~J., and {Garmire}, G.~P. (2001).
\newblock {Chandra Observations of the Pulsar Wind Nebula in Supernova Remnant
  G0.9+0.1}.
\newblock \emph{\apjl} 556, L107--L111.
\newblock \doi{10.1086/322982}
\bibAnnoteFile{Gaensler2001}

\bibitem[{{Gelfand} et~al.(2009){Gelfand}, {Slane}, and {Zhang}}]{Gelfand2009}
{Gelfand}, J.~D., {Slane}, P.~O., and {Zhang}, W. (2009).
\newblock {A Dynamical Model for the Evolution of a Pulsar Wind Nebula Inside a
  Nonradiative Supernova Remnant}.
\newblock \emph{\apj} 703, 2051--2067.
\newblock \doi{10.1088/0004-637X/703/2/2051}
\bibAnnoteFile{Gelfand2009}

\bibitem[{{Giordano} et~al.(2012){Giordano}, {Naumann-Godo}, {Ballet},
  {Bechtol}, {Funk}, {Lande} et~al.}]{2012ApJ...744L...2G}
{Giordano}, F., {Naumann-Godo}, M., {Ballet}, J., {Bechtol}, K., {Funk}, S.,
  {Lande}, J., et~al. (2012).
\newblock {Fermi Large Area Telescope Detection of the Young Supernova Remnant
  Tycho}.
\newblock \emph{\apjl} 744, L2.
\newblock \doi{10.1088/2041-8205/744/1/L2}
\bibAnnoteFile{2012ApJ...744L...2G}

\bibitem[{{Gotthelf} et~al.(2021){Gotthelf}, {Safi-Harb}, {Straal}, and
  {Gelfand}}]{Gotthelf2021}
{Gotthelf}, E.~V., {Safi-Harb}, S., {Straal}, S.~M., and {Gelfand}, J.~D.
  (2021).
\newblock {X-Ray Spectroscopy of the Highly Magnetized Pulsar PSR J1846-0258,
  Its Wind Nebula, and Hosting Supernova Remnant Kes 75}.
\newblock \emph{\apj} 908, 212.
\newblock \doi{10.3847/1538-4357/abd32b}
\bibAnnoteFile{Gotthelf2021}

\bibitem[{{Greco} et~al.(2021){Greco}, {Miceli}, {Orlando}, {Olmi}, {Bocchino},
  {Nagataki} et~al.}]{Greco2021}
{Greco}, E., {Miceli}, M., {Orlando}, S., {Olmi}, B., {Bocchino}, F.,
  {Nagataki}, S., et~al. (2021).
\newblock {Indication of a Pulsar Wind Nebula in the Hard X-Ray Emission from
  SN 1987A}.
\newblock \emph{\apjl} 908, L45.
\newblock \doi{10.3847/2041-8213/abdf5a}
\bibAnnoteFile{Greco2021}

\bibitem[{{Greco} et~al.(2022){Greco}, {Miceli}, {Orlando}, {Olmi}, {Bocchino},
  {Nagataki} et~al.}]{Greco2022}
{Greco}, E., {Miceli}, M., {Orlando}, S., {Olmi}, B., {Bocchino}, F.,
  {Nagataki}, S., et~al. (2022).
\newblock {Additional Evidence for a Pulsar Wind Nebula in the Heart of SN
  1987A from Multiepoch X-Ray Data and MHD Modeling}.
\newblock \emph{\apj} 931, 132.
\newblock \doi{10.3847/1538-4357/ac679d}
\bibAnnoteFile{Greco2022}

\bibitem[{{Grefenstette} et~al.(2017){Grefenstette}, {Fryer}, {Harrison},
  {Boggs}, {DeLaney}, {Laming} et~al.}]{2017ApJ...834...19G}
{Grefenstette}, B.~W., {Fryer}, C.~L., {Harrison}, F.~A., {Boggs}, S.~E.,
  {DeLaney}, T., {Laming}, J.~M., et~al. (2017).
\newblock {The Distribution of Radioactive $^{44}$Ti in Cassiopeia A}.
\newblock \emph{\apj} 834, 19.
\newblock \doi{10.3847/1538-4357/834/1/19}
\bibAnnoteFile{2017ApJ...834...19G}

\bibitem[{{Grefenstette} et~al.(2014){Grefenstette}, {Harrison}, {Boggs},
  {Reynolds}, {Fryer}, {Madsen} et~al.}]{2014Natur.506..339G}
{Grefenstette}, B.~W., {Harrison}, F.~A., {Boggs}, S.~E., {Reynolds}, S.~P.,
  {Fryer}, C.~L., {Madsen}, K.~K., et~al. (2014).
\newblock {Asymmetries in core-collapse supernovae from maps of radioactive
  $^{44}$Ti in CassiopeiaA}.
\newblock \emph{\nat} 506, 339--342.
\newblock \doi{10.1038/nature12997}
\bibAnnoteFile{2014Natur.506..339G}

\bibitem[{{Grefenstette} et~al.(2015){Grefenstette}, {Reynolds}, {Harrison},
  {Humensky}, {Boggs}, {Fryer} et~al.}]{NuSTAR_CasA15}
{Grefenstette}, B.~W., {Reynolds}, S.~P., {Harrison}, F.~A., {Humensky}, T.~B.,
  {Boggs}, S.~E., {Fryer}, C.~L., et~al. (2015).
\newblock {Locating the Most Energetic Electrons in Cassiopeia A}.
\newblock \emph{\apj} 802, 15.
\newblock \doi{10.1088/0004-637X/802/1/15}
\bibAnnoteFile{NuSTAR_CasA15}

\bibitem[{{Guest} et~al.(2019){Guest}, {Safi-Harb}, and {Tang}}]{Guest2019}
{Guest}, B.~T., {Safi-Harb}, S., and {Tang}, X. (2019).
\newblock {The deepest Chandra X-ray study of the plerionic supernova remnant
  G21.5-0.9}.
\newblock \emph{\mnras} 482, 1031--1042.
\newblock \doi{10.1093/mnras/sty2635}
\bibAnnoteFile{Guest2019}

\bibitem[{{H.~E.~S.~S. Collaboration} et~al.(2018){H.~E.~S.~S. Collaboration},
  {Abdalla}, {Abramowski}, {Aharonian}, {Ait Benkhali}, {Akhperjanian}
  et~al.}]{hess18}
{H.~E.~S.~S. Collaboration}, {Abdalla}, H., {Abramowski}, A., {Aharonian}, F.,
  {Ait Benkhali}, F., {Akhperjanian}, A.~G., et~al. (2018).
\newblock {The population of TeV pulsar wind nebulae in the H.E.S.S. Galactic
  Plane Survey}.
\newblock \emph{\aap} 612, A2.
\newblock \doi{10.1051/0004-6361/201629377}
\bibAnnoteFile{hess18}

\bibitem[{{H.~E.~S.~S. Collaboration} et~al.(2023){H.~E.~S.~S. Collaboration},
  {Aharonian}, {Ait Benkhali}, {Aschersleben}, {Ashkar}, {Backes}
  et~al.}]{HESS2023}
{H.~E.~S.~S. Collaboration}, {Aharonian}, F., {Ait Benkhali}, F.,
  {Aschersleben}, J., {Ashkar}, H., {Backes}, M., et~al. (2023).
\newblock {Detection of extended {\ensuremath{\gamma}}-ray emission around the
  Geminga pulsar with H.E.S.S.}
\newblock \emph{\aap} 673, A148.
\newblock \doi{10.1051/0004-6361/202245776}
\bibAnnoteFile{HESS2023}

\bibitem[{{Halpern} et~al.(2014){Halpern}, {Tomsick}, {Gotthelf}, {Camilo},
  {Ng}, {Bodaghee} et~al.}]{Halpern2014}
{Halpern}, J.~P., {Tomsick}, J.~A., {Gotthelf}, E.~V., {Camilo}, F., {Ng},
  C.~Y., {Bodaghee}, A., et~al. (2014).
\newblock {Discovery of X-Ray Pulsations from the INTEGRAL Source IGR
  J11014-6103}.
\newblock \emph{\apjl} 795, L27.
\newblock \doi{10.1088/2041-8205/795/2/L27}
\bibAnnoteFile{Halpern2014}

\bibitem[{{Hanbury Brown} and {Hazard}(1952)}]{1952Natur.170..364H}
{Hanbury Brown}, R. and {Hazard}, C. (1952).
\newblock {Radio-Frequency Radiation from Tycho Brahe's Supernova (A.D. 1572)}.
\newblock \emph{\nat} 170, 364--365.
\newblock \doi{10.1038/170364a0}
\bibAnnoteFile{1952Natur.170..364H}

\bibitem[{{Harrison} et~al.(2013){Harrison}, {Craig}, {Christensen}, {Hailey},
  {Zhang}, {Boggs} et~al.}]{Harrison2013}
{Harrison}, F.~A., {Craig}, W.~W., {Christensen}, F.~E., {Hailey}, C.~J.,
  {Zhang}, W.~W., {Boggs}, S.~E., et~al. (2013).
\newblock {The Nuclear Spectroscopic Telescope Array (NuSTAR) High-energy X-Ray
  Mission}.
\newblock \emph{\apj} 770, 103.
\newblock \doi{10.1088/0004-637X/770/2/103}
\bibAnnoteFile{Harrison2013}

\bibitem[{{Hattori} et~al.(2020){Hattori}, {Straal}, {Zhang}, {Temim},
  {Gelfand}, and {Slane}}]{Hattori2020}
{Hattori}, S., {Straal}, S.~M., {Zhang}, E., {Temim}, T., {Gelfand}, J.~D., and
  {Slane}, P.~O. (2020).
\newblock {The Nonstandard Properties of a ``Standard'' PWN: Unveiling the
  Mysteries of PWN G21.5-0.9 Using Its IR and X-Ray Emission}.
\newblock \emph{\apj} 904, 32.
\newblock \doi{10.3847/1538-4357/abba32}
\bibAnnoteFile{Hattori2020}

\bibitem[{{Helder} and {Vink}(2008)}]{2008ApJ...686.1094H}
{Helder}, E.~A. and {Vink}, J. (2008).
\newblock {Characterizing the Nonthermal Emission of Cassiopeia A}.
\newblock \emph{\apj} 686, 1094--1102.
\newblock \doi{10.1086/591242}
\bibAnnoteFile{2008ApJ...686.1094H}

\bibitem[{{Helfand} and {Becker}(1987)}]{Helfand1987}
{Helfand}, D.~J. and {Becker}, R.~H. (1987).
\newblock {G0.9+0.1 and the Emerging Class of Composite Supernova Remnants}.
\newblock \emph{\apj} 314, 203.
\newblock \doi{10.1086/165050}
\bibAnnoteFile{Helfand1987}

\bibitem[{{Hitomi Collaboration} et~al.(2018){Hitomi Collaboration},
  {Aharonian}, {Akamatsu}, {Akimoto}, {Allen}, {Angelini} et~al.}]{hitomi18}
{Hitomi Collaboration}, {Aharonian}, F., {Akamatsu}, H., {Akimoto}, F.,
  {Allen}, S.~W., {Angelini}, L., et~al. (2018).
\newblock {Hitomi X-ray observation of the pulsar wind nebula G21.5-0.9}.
\newblock \emph{\pasj} 70, 38.
\newblock \doi{10.1093/pasj/psy027}
\bibAnnoteFile{hitomi18}

\bibitem[{{Holler} et~al.(2012){Holler}, {Sch{\"o}ck}, {Eger}, {Kie{\ss}ling},
  {Valerius}, and {Stegmann}}]{Holler12}
{Holler}, M., {Sch{\"o}ck}, F.~M., {Eger}, P., {Kie{\ss}ling}, D., {Valerius},
  K., and {Stegmann}, C. (2012).
\newblock {Spatially resolved X-ray spectroscopy and modeling of the nonthermal
  emission of the pulsar wind nebula in G0.9+0.1}.
\newblock \emph{\aap} 539, A24.
\newblock \doi{10.1051/0004-6361/201118121}
\bibAnnoteFile{Holler12}

\bibitem[{{Hotokezaka} et~al.(2016){Hotokezaka}, {Wanajo}, {Tanaka}, {Bamba},
  {Terada}, and {Piran}}]{2016MNRAS.459...35H}
{Hotokezaka}, K., {Wanajo}, S., {Tanaka}, M., {Bamba}, A., {Terada}, Y., and
  {Piran}, T. (2016).
\newblock {Radioactive decay products in neutron star merger ejecta: heating
  efficiency and {\ensuremath{\gamma}}-ray emission}.
\newblock \emph{\mnras} 459, 35--43.
\newblock \doi{10.1093/mnras/stw404}
\bibAnnoteFile{2016MNRAS.459...35H}

\bibitem[{{Hwang} et~al.(2002){Hwang}, {Decourchelle}, {Holt}, and
  {Petre}}]{2002ApJ...581.1101H}
{Hwang}, U., {Decourchelle}, A., {Holt}, S.~S., and {Petre}, R. (2002).
\newblock {Thermal and Nonthermal X-Ray Emission from the Forward Shock in
  Tycho's Supernova Remnant}.
\newblock \emph{\apj} 581, 1101--1115.
\newblock \doi{10.1086/344366}
\bibAnnoteFile{2002ApJ...581.1101H}

\bibitem[{{Jansen} et~al.(2001){Jansen}, {Lumb}, {Altieri}, {Clavel}, {Ehle},
  {Erd} et~al.}]{Jansen2001}
{Jansen}, F., {Lumb}, D., {Altieri}, B., {Clavel}, J., {Ehle}, M., {Erd}, C.,
  et~al. (2001).
\newblock {XMM-Newton observatory. I. The spacecraft and operations}.
\newblock \emph{\aap} 365, L1--L6.
\newblock \doi{10.1051/0004-6361:20000036}
\bibAnnoteFile{Jansen2001}

\bibitem[{{Kargaltsev} et~al.(2017){Kargaltsev}, {Pavlov}, {Klingler}, and
  {Rangelov}}]{Kargaltsev2017}
{Kargaltsev}, O., {Pavlov}, G.~G., {Klingler}, N., and {Rangelov}, B. (2017).
\newblock {Pulsar wind nebulae created by fast-moving pulsars}.
\newblock \emph{Journal of Plasma Physics} 83, 635830501.
\newblock \doi{10.1017/S0022377817000630}
\bibAnnoteFile{Kargaltsev2017}

\bibitem[{{Kennel} and {Coroniti}(1984{\natexlab{a}})}]{kennel1984a}
{Kennel}, C.~F. and {Coroniti}, F.~V. (1984{\natexlab{a}}).
\newblock {Confinement of the Crab pulsar's wind by its supernova remnant}.
\newblock \emph{\apj} 283, 694--709.
\newblock \doi{10.1086/162356}
\bibAnnoteFile{kennel1984a}

\bibitem[{{Kennel} and {Coroniti}(1984{\natexlab{b}})}]{KC1984}
{Kennel}, C.~F. and {Coroniti}, F.~V. (1984{\natexlab{b}}).
\newblock {Magnetohydrodynamic model of Crab nebula radiation.}
\newblock \emph{\apj} 283, 710--730.
\newblock \doi{10.1086/162357}
\bibAnnoteFile{KC1984}

\bibitem[{{Klingler} et~al.(2022){Klingler}, {Hare}, {Kargaltsev}, {Pavlov},
  and {Tomsick}}]{Klinger2022}
{Klingler}, N., {Hare}, J., {Kargaltsev}, O., {Pavlov}, G.~G., and {Tomsick},
  J. (2022).
\newblock {A NuSTAR and Chandra Investigation of the Misaligned Outflow of PSR
  J1101-6101 and the Lighthouse Pulsar Wind Nebula}.
\newblock \emph{arXiv e-prints} ,
  arXiv:2212.03952\doi{10.48550/arXiv.2212.03952}
\bibAnnoteFile{Klinger2022}

\bibitem[{{Kolb} et~al.(2017){Kolb}, {Blondin}, {Slane}, and
  {Temim}}]{Kolb2017}
{Kolb}, C., {Blondin}, J., {Slane}, P., and {Temim}, T. (2017).
\newblock {Evolution of a Pulsar Wind Nebula within a Composite Supernova
  Remnant}.
\newblock \emph{\apj} 844, 1.
\newblock \doi{10.3847/1538-4357/aa75ce}
\bibAnnoteFile{Kolb2017}

\bibitem[{{Komissarov} and {Lyubarsky}(2004)}]{komissarov2004}
{Komissarov}, S.~S. and {Lyubarsky}, Y.~E. (2004).
\newblock {Synchrotron nebulae created by anisotropic magnetized pulsar winds}.
\newblock \emph{\mnras} 349, 779--792.
\newblock \doi{10.1111/j.1365-2966.2004.07597.x}
\bibAnnoteFile{komissarov2004}

\bibitem[{{Korobkin} et~al.(2020){Korobkin}, {Hungerford}, {Fryer}, {Mumpower},
  {Misch}, {Sprouse} et~al.}]{2020ApJ...889..168K}
{Korobkin}, O., {Hungerford}, A.~M., {Fryer}, C.~L., {Mumpower}, M.~R.,
  {Misch}, G.~W., {Sprouse}, T.~M., et~al. (2020).
\newblock {Gamma Rays from Kilonova: A Potential Probe of r-process
  Nucleosynthesis}.
\newblock \emph{\apj} 889, 168.
\newblock \doi{10.3847/1538-4357/ab64d8}
\bibAnnoteFile{2020ApJ...889..168K}

\bibitem[{{Krause} et~al.(2008{\natexlab{a}}){Krause}, {Birkmann}, {Usuda},
  {Hattori}, {Goto}, {Rieke} et~al.}]{2008Sci...320.1195K}
{Krause}, O., {Birkmann}, S.~M., {Usuda}, T., {Hattori}, T., {Goto}, M.,
  {Rieke}, G.~H., et~al. (2008{\natexlab{a}}).
\newblock {The Cassiopeia A Supernova Was of Type IIb}.
\newblock \emph{Science} 320, 1195.
\newblock \doi{10.1126/science.1155788}
\bibAnnoteFile{2008Sci...320.1195K}

\bibitem[{{Krause} et~al.(2008{\natexlab{b}}){Krause}, {Tanaka}, {Usuda},
  {Hattori}, {Goto}, {Birkmann} et~al.}]{2008Natur.456..617K}
{Krause}, O., {Tanaka}, M., {Usuda}, T., {Hattori}, T., {Goto}, M., {Birkmann},
  S., et~al. (2008{\natexlab{b}}).
\newblock {Tycho Brahe's 1572 supernova as a standard typeIa as revealed by its
  light-echo spectrum}.
\newblock \emph{\nat} 456, 617--619.
\newblock \doi{10.1038/nature07608}
\bibAnnoteFile{2008Natur.456..617K}

\bibitem[{{Lattimer} and {Schramm}(1974)}]{1974ApJ...192L.145L}
{Lattimer}, J.~M. and {Schramm}, D.~N. (1974).
\newblock {Black-Hole-Neutron-Star Collisions}.
\newblock \emph{\apjl} 192, L145.
\newblock \doi{10.1086/181612}
\bibAnnoteFile{1974ApJ...192L.145L}

\bibitem[{{Lopez} et~al.(2015){Lopez}, {Grefenstette}, {Reynolds}, {An},
  {Boggs}, {Christensen} et~al.}]{2015ApJ...814..132L}
{Lopez}, L.~A., {Grefenstette}, B.~W., {Reynolds}, S.~P., {An}, H., {Boggs},
  S.~E., {Christensen}, F.~E., et~al. (2015).
\newblock {A Spatially Resolved Study of the Synchrotron Emission and Titanium
  in Tycho{\textquoteright}s Supernova Remnant Using NuSTAR}.
\newblock \emph{\apj} 814, 132.
\newblock \doi{10.1088/0004-637X/814/2/132}
\bibAnnoteFile{2015ApJ...814..132L}

\bibitem[{{Madsen} et~al.(2017{\natexlab{a}}){Madsen}, {Beardmore}, {Forster},
  {Guainazzi}, {Marshall}, {Miller} et~al.}]{Madsen2017}
{Madsen}, K.~K., {Beardmore}, A.~P., {Forster}, K., {Guainazzi}, M.,
  {Marshall}, H.~L., {Miller}, E.~D., et~al. (2017{\natexlab{a}}).
\newblock {IACHEC Cross-calibration of Chandra, NuSTAR, Swift, Suzaku,
  XMM-Newton with 3C 273 and PKS 2155-304}.
\newblock \emph{\aj} 153, 2.
\newblock \doi{10.3847/1538-3881/153/1/2}
\bibAnnoteFile{Madsen2017}

\bibitem[{{Madsen} et~al.(2017{\natexlab{b}}){Madsen}, {Christensen}, {Craig},
  {Forster}, {Grefenstette}, {Harrison} et~al.}]{Madsen2017b}
{Madsen}, K.~K., {Christensen}, F.~E., {Craig}, W.~W., {Forster}, K.~W.,
  {Grefenstette}, B.~W., {Harrison}, F.~A., et~al. (2017{\natexlab{b}}).
\newblock {Observational Artifacts of NuSTAR: Ghost Rays and Stray Light}.
\newblock \emph{arXiv e-prints} ,
  arXiv:1711.02719\doi{10.48550/arXiv.1711.02719}
\bibAnnoteFile{Madsen2017b}

\bibitem[{{Madsen} et~al.(2020){Madsen}, {Fryer}, {Grefenstette}, {Lopez},
  {Reynolds}, and {Zoglauer}}]{Madsen2020}
{Madsen}, K.~K., {Fryer}, C.~L., {Grefenstette}, B.~W., {Lopez}, L.~A.,
  {Reynolds}, S., and {Zoglauer}, A. (2020).
\newblock {NuSTAR Observations of G11.2-0.3}.
\newblock \emph{\apj} 889, 23.
\newblock \doi{10.3847/1538-4357/ab54ca}
\bibAnnoteFile{Madsen2020}

\bibitem[{{Madsen} et~al.(2015{\natexlab{a}}){Madsen}, {Reynolds}, {Harrison},
  {An}, {Boggs}, {Christensen} et~al.}]{Madsen2015}
{Madsen}, K.~K., {Reynolds}, S., {Harrison}, F., {An}, H., {Boggs}, S.,
  {Christensen}, F.~E., et~al. (2015{\natexlab{a}}).
\newblock {Broadband X-ray Imaging and Spectroscopy of the Crab Nebula and
  Pulsar with NuSTAR}.
\newblock \emph{\apj} 801, 66.
\newblock \doi{10.1088/0004-637X/801/1/66}
\bibAnnoteFile{Madsen2015}

\bibitem[{{Madsen} et~al.(2015{\natexlab{b}}){Madsen}, {Reynolds}, {Harrison},
  {An}, {Boggs}, {Christensen} et~al.}]{Madsen2015b}
{Madsen}, K.~K., {Reynolds}, S., {Harrison}, F., {An}, H., {Boggs}, S.,
  {Christensen}, F.~E., et~al. (2015{\natexlab{b}}).
\newblock {Broadband X-ray Imaging and Spectroscopy of the Crab Nebula and
  Pulsar with NuSTAR}.
\newblock \emph{\apj} 801, 66.
\newblock \doi{10.1088/0004-637X/801/1/66}
\bibAnnoteFile{Madsen2015b}

\bibitem[{{Maeda} et~al.(2010){Maeda}, {R{\"o}pke}, {Fink}, {Hillebrandt},
  {Travaglio}, and {Thielemann}}]{maeda10}
{Maeda}, K., {R{\"o}pke}, F.~K., {Fink}, M., {Hillebrandt}, W., {Travaglio},
  C., and {Thielemann}, F.~K. (2010).
\newblock {Nucleosynthesis in Two-Dimensional Delayed Detonation Models of Type
  Ia Supernova Explosions}.
\newblock \emph{\apj} 712, 624--638.
\newblock \doi{10.1088/0004-637X/712/1/624}
\bibAnnoteFile{maeda10}

\bibitem[{{Maeda} et~al.(2009){Maeda}, {Uchiyama}, {Bamba}, {Kosugi},
  {Tsunemi}, {Helder} et~al.}]{2009PASJ...61.1217M}
{Maeda}, Y., {Uchiyama}, Y., {Bamba}, A., {Kosugi}, H., {Tsunemi}, H.,
  {Helder}, E.~A., et~al. (2009).
\newblock {Suzaku X-Ray Imaging and Spectroscopy of Cassiopeia A}.
\newblock \emph{\pasj} 61, 1217.
\newblock \doi{10.1093/pasj/61.6.1217}
\bibAnnoteFile{2009PASJ...61.1217M}

\bibitem[{{Magkotsios} et~al.(2010){Magkotsios}, {Timmes}, {Hungerford},
  {Fryer}, {Young}, and {Wiescher}}]{2010ApJS..191...66M}
{Magkotsios}, G., {Timmes}, F.~X., {Hungerford}, A.~L., {Fryer}, C.~L.,
  {Young}, P.~A., and {Wiescher}, M. (2010).
\newblock {Trends in $^{44}$Ti and $^{56}$Ni from Core-collapse Supernovae}.
\newblock \emph{\apjs} 191, 66--95.
\newblock \doi{10.1088/0067-0049/191/1/66}
\bibAnnoteFile{2010ApJS..191...66M}

\bibitem[{{Manconi} et~al.(2020){Manconi}, {Di Mauro}, and
  {Donato}}]{Manconi2020}
{Manconi}, S., {Di Mauro}, M., and {Donato}, F. (2020).
\newblock {Contribution of pulsars to cosmic-ray positrons in light of recent
  observation of inverse-Compton halos}.
\newblock \emph{\prd} 102, 023015.
\newblock \doi{10.1103/PhysRevD.102.023015}
\bibAnnoteFile{Manconi2020}

\bibitem[{{Mori} et~al.(2021){Mori}, {An}, {Burgess}, {Capasso}, {Dingus},
  {Gelfand} et~al.}]{Mori2021}
{Mori}, K., {An}, H., {Burgess}, D., {Capasso}, M., {Dingus}, B., {Gelfand},
  J., et~al. (2021).
\newblock {NuSTAR broad-band X-ray observational campaign of energetic pulsar
  wind nebulae in synergy with VERITAS, HAWC and Fermi gamma-ray telescopes}.
\newblock \emph{arXiv e-prints} , arXiv:2108.00557
\bibAnnoteFile{Mori2021}

\bibitem[{{Mori} et~al.(2004){Mori}, {Burrows}, {Hester}, {Pavlov}, {Shibata},
  and {Tsunemi}}]{Mori2004}
{Mori}, K., {Burrows}, D.~N., {Hester}, J.~J., {Pavlov}, G.~G., {Shibata}, S.,
  and {Tsunemi}, H. (2004).
\newblock {Spatial Variation of the X-Ray Spectrum of the Crab Nebula}.
\newblock \emph{\apj} 609, 186--193.
\newblock \doi{10.1086/421011}
\bibAnnoteFile{Mori2004}

\bibitem[{{Ng} and {Romani}(2004)}]{ng04}
{Ng}, C.~Y. and {Romani}, R.~W. (2004).
\newblock {Fitting Pulsar Wind Tori}.
\newblock \emph{ApJ} 601, 479--484.
\newblock \doi{10.1086/380486}
\bibAnnoteFile{ng04}

\bibitem[{{Nynka} et~al.(2014){Nynka}, {Hailey}, {Reynolds}, {An}, {Baganoff},
  {Boggs} et~al.}]{Nynka2014}
{Nynka}, M., {Hailey}, C.~J., {Reynolds}, S.~P., {An}, H., {Baganoff}, F.~K.,
  {Boggs}, S.~E., et~al. (2014).
\newblock {NuSTAR Study of Hard X-Ray Morphology and Spectroscopy of PWN
  G21.5-0.9}.
\newblock \emph{\apj} 789, 72.
\newblock \doi{10.1088/0004-637X/789/1/72}
\bibAnnoteFile{Nynka2014}

\bibitem[{{Okuno} et~al.(2020){Okuno}, {Tanaka}, {Uchida}, {Aharonian},
  {Uchiyama}, {Tsuru} et~al.}]{2020ApJ...894...50O}
{Okuno}, T., {Tanaka}, T., {Uchida}, H., {Aharonian}, F.~A., {Uchiyama}, Y.,
  {Tsuru}, T.~G., et~al. (2020).
\newblock {Time Variability of Nonthermal X-Ray Stripes in Tycho's Supernova
  Remnant with Chandra}.
\newblock \emph{\apj} 894, 50.
\newblock \doi{10.3847/1538-4357/ab837e}
\bibAnnoteFile{2020ApJ...894...50O}

\bibitem[{{Olmi} and {Bucciantini}(2019)}]{Olmi2019}
{Olmi}, B. and {Bucciantini}, N. (2019).
\newblock {On the origin of jet-like features in bow shock pulsar wind
  nebulae}.
\newblock \emph{\mnras} 490, 3608--3615.
\newblock \doi{10.1093/mnras/stz2819}
\bibAnnoteFile{Olmi2019}

\bibitem[{{Olmi} and {Bucciantini}(2023)}]{Olmi2023}
{Olmi}, B. and {Bucciantini}, N. (2023).
\newblock {From young to old: The evolutionary path of Pulsar Wind Nebulae}.
\newblock \emph{\pasa} 40, e007.
\newblock \doi{10.1017/pasa.2023.5}
\bibAnnoteFile{Olmi2023}

\bibitem[{{Orlando} et~al.(2022){Orlando}, {Wongwathanarat}, {Janka}, {Miceli},
  {Nagataki}, {Ono} et~al.}]{2022A&A...666A...2O}
{Orlando}, S., {Wongwathanarat}, A., {Janka}, H.~T., {Miceli}, M., {Nagataki},
  S., {Ono}, M., et~al. (2022).
\newblock {Evidence for past interaction with an asymmetric circumstellar shell
  in the young SNR Cassiopeia A}.
\newblock \emph{\aap} 666, A2.
\newblock \doi{10.1051/0004-6361/202243258}
\bibAnnoteFile{2022A&A...666A...2O}

\bibitem[{{Parizot} et~al.(2006){Parizot}, {Marcowith}, {Ballet}, and
  {Gallant}}]{parizot06}
{Parizot}, E., {Marcowith}, A., {Ballet}, J., and {Gallant}, Y.~A. (2006).
\newblock {Observational constraints on energetic particle diffusion in young
  supernovae remnants: amplified magnetic field and maximum energy}.
\newblock \emph{\aap} 453, 387--395.
\newblock \doi{10.1051/0004-6361:20064985}
\bibAnnoteFile{parizot06}

\bibitem[{{Park} et~al.(2023{\natexlab{a}}){Park}, {Kim}, {Woo}, {An}, {Mori},
  {Reynolds} et~al.}]{Park2023a}
{Park}, J., {Kim}, C., {Woo}, J., {An}, H., {Mori}, K., {Reynolds}, S.~P.,
  et~al. (2023{\natexlab{a}}).
\newblock {A Broadband X-Ray Study of the Rabbit Pulsar Wind Nebula Powered by
  PSR J1418-6058}.
\newblock \emph{\apj} 945, 66.
\newblock \doi{10.3847/1538-4357/acba0e}
\bibAnnoteFile{Park2023a}

\bibitem[{{Park} et~al.(2023{\natexlab{b}}){Park}, {Kim}, {Woo}, {An}, {Mori},
  {Reynolds} et~al.}]{Park2023b}
{Park}, J., {Kim}, C., {Woo}, J., {An}, H., {Mori}, K., {Reynolds}, S.~P.,
  et~al. (2023{\natexlab{b}}).
\newblock {X-Ray Studies of the Pulsar PSR J1420-6048 and Its TeV Pulsar Wind
  Nebula in the Kookaburra Region}.
\newblock \emph{\apj} 945, 33.
\newblock \doi{10.3847/1538-4357/acb1b0}
\bibAnnoteFile{Park2023b}

\bibitem[{{Pavan} et~al.(2016){Pavan}, {P{\"u}hlhofer}, {Bordas}, {Audard},
  {Balbo}, {Bozzo} et~al.}]{Pavan2016}
{Pavan}, L., {P{\"u}hlhofer}, G., {Bordas}, P., {Audard}, M., {Balbo}, M.,
  {Bozzo}, E., et~al. (2016).
\newblock {Closer view of the IGR J11014-6103 outflows}.
\newblock \emph{\aap} 591, A91.
\newblock \doi{10.1051/0004-6361/201527703}
\bibAnnoteFile{Pavan2016}

\bibitem[{{Porquet} et~al.(2003){Porquet}, {Decourchelle}, and
  {Warwick}}]{Porquet2003}
{Porquet}, D., {Decourchelle}, A., and {Warwick}, R.~S. (2003).
\newblock {XMM-Newton spectral analysis of the Pulsar Wind Nebula within the
  composite SNR <ASTROBJ>G0.9+0.1</ASTROBJ>}.
\newblock \emph{\aap} 401, 197--203.
\newblock \doi{10.1051/0004-6361:20021670}
\bibAnnoteFile{Porquet2003}

\bibitem[{{Porter} et~al.(2022){Porter}, {J{\'o}hannesson}, and
  {Moskalenko}}]{Porter2022}
{Porter}, T.~A., {J{\'o}hannesson}, G., and {Moskalenko}, I.~V. (2022).
\newblock {The GALPROP Cosmic-ray Propagation and Nonthermal Emissions
  Framework: Release v57}.
\newblock \emph{\apjs} 262, 30.
\newblock \doi{10.3847/1538-4365/ac80f6}
\bibAnnoteFile{Porter2022}

\bibitem[{{Pravdo} and {Smith}(1979)}]{1979ApJ...234L.195P}
{Pravdo}, S.~H. and {Smith}, B.~W. (1979).
\newblock {X-ray evidence for electron-ion equilibrium and ionization
  nonequilibrium in Young supernova remnants}.
\newblock \emph{\apjl} 234, L195--L198.
\newblock \doi{10.1086/183138}
\bibAnnoteFile{1979ApJ...234L.195P}

\bibitem[{{Renaud} et~al.(2010){Renaud}, {Marandon}, {Gotthelf}, {Rodriguez},
  {Terrier}, {Mattana} et~al.}]{renaud10}
{Renaud}, M., {Marandon}, V., {Gotthelf}, E.~V., {Rodriguez}, J., {Terrier},
  R., {Mattana}, F., et~al. (2010).
\newblock {Discovery of a Highly Energetic Pulsar Associated with IGR
  J14003-6326 in the Young Uncataloged Galactic Supernova Remnant G310.6-1.6}.
\newblock \emph{\apj} 716, 663--670.
\newblock \doi{10.1088/0004-637X/716/1/663}
\bibAnnoteFile{renaud10}

\bibitem[{{Reynolds}(2009)}]{Reynolds09}
{Reynolds}, S.~P. (2009).
\newblock {Synchrotron-Loss Spectral Breaks in Pulsar-Wind Nebulae and
  Extragalactic Jets}.
\newblock \emph{\apj} 703, 662--670.
\newblock \doi{10.1088/0004-637X/703/1/662}
\bibAnnoteFile{Reynolds09}

\bibitem[{{Reynolds}(2016)}]{Reynolds2016}
{Reynolds}, S.~P. (2016).
\newblock {Hard X-ray emission from pulsar-wind nebulae}.
\newblock \emph{Journal of Plasma Physics} 82, 635820501.
\newblock \doi{10.1017/S0022377816000751}
\bibAnnoteFile{Reynolds2016}

\bibitem[{{Reynolds} et~al.(2008){Reynolds}, {Borkowski}, {Green}, {Hwang},
  {Harrus}, and {Petre}}]{reynolds08}
{Reynolds}, S.~P., {Borkowski}, K.~J., {Green}, D.~A., {Hwang}, U., {Harrus},
  I., and {Petre}, R. (2008).
\newblock {The Youngest Galactic Supernova Remnant: G1.9+0.3}.
\newblock \emph{\apjl} 680, L41.
\newblock \doi{10.1086/589570}
\bibAnnoteFile{reynolds08}

\bibitem[{{Reynolds} et~al.(2009){Reynolds}, {Borkowski}, {Green}, {Hwang},
  {Harrus}, and {Petre}}]{reynolds09b}
{Reynolds}, S.~P., {Borkowski}, K.~J., {Green}, D.~A., {Hwang}, U., {Harrus},
  I., and {Petre}, R. (2009).
\newblock {X-Ray Spectral Variations in the Youngest Galactic Supernova Remnant
  G1.9+0.3}.
\newblock \emph{\apjl} 695, L149--L153.
\newblock \doi{10.1088/0004-637X/695/2/L149}
\bibAnnoteFile{reynolds09b}

\bibitem[{{Reynolds} et~al.(2018){Reynolds}, {Borkowski}, and
  {Gwynne}}]{reynolds18}
{Reynolds}, S.~P., {Borkowski}, K.~J., and {Gwynne}, P.~H. (2018).
\newblock {Expansion and Brightness Changes in the Pulsar-wind Nebula in the
  Composite Supernova Remnant Kes 75}.
\newblock \emph{\apj} 856, 133.
\newblock \doi{10.3847/1538-4357/aab3d3}
\bibAnnoteFile{reynolds18}

\bibitem[{{Reynolds} and {Chevalier}(1984)}]{Reynolds1984}
{Reynolds}, S.~P. and {Chevalier}, R.~A. (1984).
\newblock {Evolution of pulsar-driven supernova remnants.}
\newblock \emph{\apj} 278, 630--648.
\newblock \doi{10.1086/161831}
\bibAnnoteFile{Reynolds1984}

\bibitem[{{Reynoso} et~al.(1997){Reynoso}, {Moffett}, {Goss}, {Dubner},
  {Dickel}, {Reynolds} et~al.}]{1997ApJ...491..816R}
{Reynoso}, E.~M., {Moffett}, D.~A., {Goss}, W.~M., {Dubner}, G.~M., {Dickel},
  J.~R., {Reynolds}, S.~P., et~al. (1997).
\newblock {A VLA Study of the Expansion of Tycho's Supernova Remnant}.
\newblock \emph{\apj} 491, 816--828.
\newblock \doi{10.1086/304997}
\bibAnnoteFile{1997ApJ...491..816R}

\bibitem[{{Sato} et~al.(2018){Sato}, {Katsuda}, {Morii}, {Bamba}, {Hughes},
  {Maeda} et~al.}]{Chandra_CasA}
{Sato}, T., {Katsuda}, S., {Morii}, M., {Bamba}, A., {Hughes}, J.~P., {Maeda},
  Y., et~al. (2018).
\newblock {X-Ray Measurements of the Particle Acceleration Properties at Inward
  Shocks in Cassiopeia A}.
\newblock \emph{\apj} 853, 46.
\newblock \doi{10.3847/1538-4357/aaa021}
\bibAnnoteFile{Chandra_CasA}

\bibitem[{{Sironi} and {Spitkovsky}(2009)}]{sironi2009}
{Sironi}, L. and {Spitkovsky}, A. (2009).
\newblock {Particle Acceleration in Relativistic Magnetized Collisionless Pair
  Shocks: Dependence of Shock Acceleration on Magnetic Obliquity}.
\newblock \emph{\apj} 698, 1523--1549.
\newblock \doi{10.1088/0004-637X/698/2/1523}
\bibAnnoteFile{sironi2009}

\bibitem[{{Sironi} and {Spitkovsky}(2014)}]{Sironi2014}
{Sironi}, L. and {Spitkovsky}, A. (2014).
\newblock {Relativistic Reconnection: An Efficient Source of Non-thermal
  Particles}.
\newblock \emph{\apjl} 783, L21.
\newblock \doi{10.1088/2041-8205/783/1/L21}
\bibAnnoteFile{Sironi2014}

\bibitem[{{Sudoh} et~al.(2019){Sudoh}, {Linden}, and {Beacom}}]{sudoh19}
{Sudoh}, T., {Linden}, T., and {Beacom}, J.~F. (2019).
\newblock {TeV halos are everywhere: Prospects for new discoveries}.
\newblock \emph{\prd} 100, 043016.
\newblock \doi{10.1103/PhysRevD.100.043016}
\bibAnnoteFile{sudoh19}

\bibitem[{{Tanaka} and {Takahara}(2011)}]{Tanaka2011}
{Tanaka}, S.~J. and {Takahara}, F. (2011).
\newblock {Study of Four Young TeV Pulsar Wind Nebulae with a Spectral
  Evolution Model}.
\newblock \emph{\apj} 741, 40.
\newblock \doi{10.1088/0004-637X/741/1/40}
\bibAnnoteFile{Tanaka2011}

\bibitem[{{Tanvir} et~al.(2013){Tanvir}, {Levan}, {Fruchter}, {Hjorth},
  {Hounsell}, {Wiersema} et~al.}]{2013Natur.500..547T}
{Tanvir}, N.~R., {Levan}, A.~J., {Fruchter}, A.~S., {Hjorth}, J., {Hounsell},
  R.~A., {Wiersema}, K., et~al. (2013).
\newblock {A `kilonova' associated with the short-duration
  {\ensuremath{\gamma}}-ray burst GRB 130603B}.
\newblock \emph{\nat} 500, 547--549.
\newblock \doi{10.1038/nature12505}
\bibAnnoteFile{2013Natur.500..547T}

\bibitem[{{Terada} et~al.(2022){Terada}, {Miwa}, {Ohsumi}, {Fujimoto},
  {Katsuda}, {Bamba} et~al.}]{2022ApJ...933..111T}
{Terada}, Y., {Miwa}, Y., {Ohsumi}, H., {Fujimoto}, S.-i., {Katsuda}, S.,
  {Bamba}, A., et~al. (2022).
\newblock {Gamma-Ray Diagnostics of r-process Nucleosynthesis in the Remnants
  of Galactic Binary Neutron-star Mergers}.
\newblock \emph{\apj} 933, 111.
\newblock \doi{10.3847/1538-4357/ac721f}
\bibAnnoteFile{2022ApJ...933..111T}

\bibitem[{{Torres} et~al.(2014){Torres}, {Cillis}, {Mart{\'\i}n}, and {de
  O{\~n}a Wilhelmi}}]{Torres2014}
{Torres}, D.~F., {Cillis}, A., {Mart{\'\i}n}, J., and {de O{\~n}a Wilhelmi}, E.
  (2014).
\newblock {Time-dependent modeling of TeV-detected, young pulsar wind nebulae}.
\newblock \emph{Journal of High Energy Astrophysics} 1, 31--62.
\newblock \doi{10.1016/j.jheap.2014.02.001}
\bibAnnoteFile{Torres2014}

\bibitem[{{Uchiyama} and {Aharonian}(2008)}]{2008ApJ...677L.105U}
{Uchiyama}, Y. and {Aharonian}, F.~A. (2008).
\newblock {Fast Variability of Nonthermal X-Ray Emission in Cassiopeia A:
  Probing Electron Acceleration in Reverse-Shocked Ejecta}.
\newblock \emph{\apjl} 677, L105.
\newblock \doi{10.1086/588190}
\bibAnnoteFile{2008ApJ...677L.105U}

\bibitem[{{Van Etten} and Romani(2011)}]{Van_Etten_2011}
{Van Etten}, A. and Romani, R.~W. (2011).
\newblock {MULTI}-{ZONE} {MODELING} {OF} {THE} {PULSAR} {WIND} {NEBULA} {HESS}
  j1825{\textendash}137.
\newblock \emph{The Astrophysical Journal} 742, 62.
\newblock \doi{10.1088/0004-637x/742/2/62}
\bibAnnoteFile{Van_Etten_2011}

\bibitem[{{Vance} et~al.(2020){Vance}, {Young}, {Fryer}, and
  {Ellinger}}]{2020ApJ...895...82V}
{Vance}, G.~S., {Young}, P.~A., {Fryer}, C.~L., and {Ellinger}, C.~I. (2020).
\newblock {Titanium and Iron in the Cassiopeia A Supernova Remnant}.
\newblock \emph{\apj} 895, 82.
\newblock \doi{10.3847/1538-4357/ab8ade}
\bibAnnoteFile{2020ApJ...895...82V}

\bibitem[{{Vink}(2012)}]{vink12}
{Vink}, J. (2012).
\newblock {Supernova remnants: the X-ray perspective}.
\newblock \emph{\aapr} 20, 49.
\newblock \doi{10.1007/s00159-011-0049-1}
\bibAnnoteFile{vink12}

\bibitem[{{Vink} et~al.(2000){Vink}, {Kaastra}, {Bleeker}, and
  {Bloemen}}]{vink00}
{Vink}, J., {Kaastra}, J.~S., {Bleeker}, J. A.~M., and {Bloemen}, H. (2000).
\newblock {The Hard X-Ray Emission and 44ti Emission Of Cas A}.
\newblock \emph{Advances in Space Research} 25, 689--694.
\newblock \doi{10.1016/S0273-1177(99)00823-6}
\bibAnnoteFile{vink00}

\bibitem[{{Vink} et~al.(2022){Vink}, {Prokhorov}, {Ferrazzoli}, {Slane},
  {Zhou}, {Asakura} et~al.}]{2022ApJ...938...40V}
{Vink}, J., {Prokhorov}, D., {Ferrazzoli}, R., {Slane}, P., {Zhou}, P.,
  {Asakura}, K., et~al. (2022).
\newblock {X-Ray Polarization Detection of Cassiopeia A with IXPE}.
\newblock \emph{\apj} 938, 40.
\newblock \doi{10.3847/1538-4357/ac8b7b}
\bibAnnoteFile{2022ApJ...938...40V}

\bibitem[{{Wang} et~al.(2020){Wang}, {N3AS Collaboration}, {Vassh}, {FIRE
  Collaboration}, {Sprouse}, {Mumpower} et~al.}]{2020ApJ...903L...3W}
{Wang}, X., {N3AS Collaboration}, {Vassh}, N., {FIRE Collaboration}, {Sprouse},
  T., {Mumpower}, M., et~al. (2020).
\newblock {MeV Gamma Rays from Fission: A Distinct Signature of Actinide
  Production in Neutron Star Mergers}.
\newblock \emph{\apjl} 903, L3.
\newblock \doi{10.3847/2041-8213/abbe18}
\bibAnnoteFile{2020ApJ...903L...3W}

\bibitem[{{Warren} et~al.(2005){Warren}, {Hughes}, {Badenes}, {Ghavamian},
  {McKee}, {Moffett} et~al.}]{2005ApJ...634..376W}
{Warren}, J.~S., {Hughes}, J.~P., {Badenes}, C., {Ghavamian}, P., {McKee},
  C.~F., {Moffett}, D., et~al. (2005).
\newblock {Cosmic-Ray Acceleration at the Forward Shock in Tycho's Supernova
  Remnant: Evidence from Chandra X-Ray Observations}.
\newblock \emph{\apj} 634, 376--389.
\newblock \doi{10.1086/496941}
\bibAnnoteFile{2005ApJ...634..376W}

\bibitem[{{Wik} et~al.(2014){Wik}, {Hornstrup}, {Molendi}, {Madejski},
  {Harrison}, and {Zoglauer}}]{Wik2014}
{Wik}, D.~R., {Hornstrup}, A., {Molendi}, S., {Madejski}, G., {Harrison},
  F.~A., and {Zoglauer}, A. (2014).
\newblock {NuSTAR Observations of the Bullet Cluster: Constraints on Inverse
  Compton Emission}.
\newblock \emph{\apj} 792, 48.
\newblock \doi{10.1088/0004-637X/792/1/48}
\bibAnnoteFile{Wik2014}

\bibitem[{{Wongwathanarat} et~al.(2017){Wongwathanarat}, {Janka}, {M{\"u}ller},
  {Pllumbi}, and {Wanajo}}]{2017ApJ...842...13W}
{Wongwathanarat}, A., {Janka}, H.-T., {M{\"u}ller}, E., {Pllumbi}, E., and
  {Wanajo}, S. (2017).
\newblock {Production and Distribution of $^{44}$Ti and $^{56}$Ni in a
  Three-dimensional Supernova Model Resembling Cassiopeia A}.
\newblock \emph{\apj} 842, 13.
\newblock \doi{10.3847/1538-4357/aa72de}
\bibAnnoteFile{2017ApJ...842...13W}

\bibitem[{{Woo} et~al.(2023){Woo}, {An}, {Gelfand}, {Hailey}, {Mori},
  {Mukherjee} et~al.}]{Woo2023}
{Woo}, J., {An}, H., {Gelfand}, J.~D., {Hailey}, C.~J., {Mori}, K.,
  {Mukherjee}, R., et~al. (2023).
\newblock {Hard X-ray observation and multiwavelength study of the PeVatron
  candidate pulsar wind nebula ``Dragonfly''}.
\newblock \emph{arXiv e-prints} ,
  arXiv:2306.07347\doi{10.48550/arXiv.2306.07347}
\bibAnnoteFile{Woo2023}

\bibitem[{{Wu} et~al.(2019){Wu}, {Banerjee}, {Metzger}, {Mart{\'\i}nez-Pinedo},
  {Aramaki}, {Burns} et~al.}]{2019ApJ...880...23W}
{Wu}, M.-R., {Banerjee}, P., {Metzger}, B.~D., {Mart{\'\i}nez-Pinedo}, G.,
  {Aramaki}, T., {Burns}, E., et~al. (2019).
\newblock {Finding the Remnants of the Milky Way's Last Neutron Star Mergers}.
\newblock \emph{\apj} 880, 23.
\newblock \doi{10.3847/1538-4357/ab2593}
\bibAnnoteFile{2019ApJ...880...23W}

\bibitem[{{Xi} et~al.(2019){Xi}, {Liu}, {Huang}, {Fang}, and {Wang}}]{Xi2019}
{Xi}, S.-Q., {Liu}, R.-Y., {Huang}, Z.-Q., {Fang}, K., and {Wang}, X.-Y.
  (2019).
\newblock {GeV Observations of the Extended Pulsar Wind Nebulae Constrain the
  Pulsar Interpretations of the Cosmic-Ray Positron Excess}.
\newblock \emph{\apj} 878, 104.
\newblock \doi{10.3847/1538-4357/ab20c9}
\bibAnnoteFile{Xi2019}

\bibitem[{{Y{\"u}ksel} et~al.(2009){Y{\"u}ksel}, {Kistler}, and
  {Stanev}}]{Yuksel2009}
{Y{\"u}ksel}, H., {Kistler}, M.~D., and {Stanev}, T. (2009).
\newblock {TeV Gamma Rays from Geminga and the Origin of the GeV Positron
  Excess}.
\newblock \emph{\prl} 103, 051101.
\newblock \doi{10.1103/PhysRevLett.103.051101}
\bibAnnoteFile{Yuksel2009}

\bibitem[{{Zoglauer} et~al.(2015){Zoglauer}, {Reynolds}, {An}, {Boggs},
  {Christensen}, {Craig} et~al.}]{zoglauer15}
{Zoglauer}, A., {Reynolds}, S.~P., {An}, H., {Boggs}, S.~E., {Christensen},
  F.~E., {Craig}, W.~W., et~al. (2015).
\newblock {The Hard X-Ray View of the Young Supernova Remnant G1.9+0.3}.
\newblock \emph{\apj} 798, 98.
\newblock \doi{10.1088/0004-637X/798/2/98}
\bibAnnoteFile{zoglauer15}

\end{thebibliography}

\end{document}